\documentclass[showpacs,amsmath,amssymb,aps,twocolumn,notitlepage]{revtex4-1}
\usepackage{bbm}
\usepackage{braket}
\usepackage{amsfonts}
\usepackage{amsmath}
\usepackage{amsthm}
\usepackage{txfonts}
\usepackage{bm}
\usepackage{graphicx}
\usepackage{color}

\usepackage{mathrsfs}

\usepackage{subfigure}
\usepackage{wrapfig}
\usepackage{graphicx}

\usepackage{amsmath}

\newcommand{\eq}[1]{Eq.~(\ref{#1})}
\newcommand{\be}{\begin{equation}}
\newcommand{\ee}{\end{equation}}
\newcommand{\overbar}[1]{\mkern 1.5mu\overline{\mkern-1.5mu#1\mkern-1.5mu}\mkern 1.5mu}

\makeatletter
\def\munderbar#1{\underline{\sbox\tw@{$#1$}\dp\tw@\z@\box\tw@}}
\makeatother

\makeatletter
\def\mbar#1{\overline{\sbox\tw@{$#1$}\dp\tw@\z@\box\tw@}}
\makeatother

\begin{document}

\title{Finite-key analysis for quantum key distribution with weak coherent pulses \\ based on Bernoulli sampling}
\author{
Shun Kawakami$^{1,2}$, Toshihiko Sasaki$^1$ and Masato Koashi$^{1,2}$}

\affiliation{
$^1$Photon Science Center, Graduate School of Engineering, The University of Tokyo, 
2-11-16 Yayoi, Bunkyo-ku, Tokyo 113-8656, Japan\\
$^2$ Department of Applied Physics, 
Graduate School of Engineering, 
The University of Tokyo, 7-3-1 Hongo, Bunkyo-ku, Tokyo 113-8656, Japan 
}

\begin{abstract}
An essential step in quantum key distribution is the estimation of parameters related to the 
leaked amount of information, which is usually done by sampling of the communication data. 
When the data size is finite, the final key rate depends on how the estimation process handles 
statistical fluctuations. 
Many of the present security analyses are based on the method with simple random sampling, 
where hypergeometric distribution or its known bounds are used for the estimation. 
Here we propose a concise method based on Bernoulli sampling, 
which is related to binomial distribution. 
Our method is suitable for the BB84 protocol with weak coherent pulses, 
reducing the number of estimated parameters to 
achieve a higher key generation rate compared to the method with simple random sampling. 
We also applied the method to prove the security of the differential-quadrature-phase-shift (DQPS) protocol 
in the finite-key regime. 
The result indicates that, 
the advantage of the DQPS protocol over the phase-encoding BB84 protocol in terms of the key rate, 
which was previously confirmed in the asymptotic regime, 
persists in the finite-key regime.

%the DQPS protocol has higher key rate than the phase-encoding BB84 protocol in the finite-key regime. 

\end{abstract}


\maketitle

\section{Introduction}

Quantum key distribution (QKD) allows two distant parties to share a secret key and 
realizes a communication with information-theoretic security by combining it with 
one-time-pad encryption. 
Since the BB84 protocol was proposed by Bennett and Brassard~\cite{1984Bennett}, 
a large number of researches on QKD have been conducted from both aspects of theory and implementations. 
The security of QKD is based only on principles of quantum physics where 
eavesdropped information is bounded from the observed parameters in a protocol. 
In practice, the estimation of this bound should take into account statistical fluctuations 
due to the finite size of communication data, 
which requires so-called finite-key analysis.  
Following the security definition with composability \cite{2009Benor,2009Renner}, 
the finite-key analyses for various QKD protocols including the BB84 protocol 
were conducted assuming 
the adversary's general 
attacks~\cite{2012Tomamichel,2012Furrer,2014Wen,2014Curty,2014Hayashi,2015Lucamarini,2015Mizutani}.

%In several recent works related to a finite-key analysis, 
%the protocol with threshold of data size is used \cite{2012Tomamichel,2014Wen,2014Curty}, in which  
%the protocol runs until 
%the data size of $Z$ basis $n_Z$ and $X$ basis $n_X$ reach some threshold 
%$n_Z^{\rm th}$ and $n_X^{\rm th}$, respectively. 
%In such a threshold protocol, 
%a basis choice is publicly disclosed at each round of protocol to trace the data size, 
%which is called iterative sifting. 
%On the other hand, a recent work \cite{2016Pfister} shows that the threshold protocol with iterative sifting is 
%not secure if we use the finite-key analysis with simple random sampling, 
%which was a conventional method to prove the security of the BB84 protocol. 
%To solve this problem, two approaches are proposed. 
%The first one \cite{2016Pfister} is changing an operation of the protocol, in which 
%a basis is disclosed after all rounds of the quantum communication. 
%Since the use of the threshold protocol is assumed in this approach, 
% additional rounds are required to 
%reduce the probability that the protocol aborts 
%due to a final data size falling the predetermined threshold. 
%The second approach \cite{2016Tamaki} is to alter the way of analysis, in which Azuma's inequality is used 
%instead of random sampling theory. 
%On the other hand, a finite-key analysis with Azuma's inequality typically 
%leads to worse bound compared to that with random sampling theory 
%because it covers Eve's attack which is optimized on disclosed information at the previous round. 

For a finite-key analysis, a simple method with a smaller number of estimation processes is preferred 
because it leads not only to a more concise security proof but also to a higher key-generation rate 
especially when the number of communication rounds is limited and statistical fluctuations are large. 
A number of recent finite-key analyses are based on the method with simple random sampling, 
which is used to model $n_1$ draws, without replacement, 
from a finite population of size $n_2$ that contains $k_2$ errors. 
The probability that the number of errors in the sample is $k_1$ obeys hypergeometric distribution 
\begin{align}
{\rm HG}(k_1;n_1,k_2,n_2):=
\frac{\binom{k_2}{k_1}\binom{n_2-k_2}{n_1-k_1}}
{\binom{n_2}{n_1}}.
\end{align} 
%For simple random sampling, 
%the size of sample and the population are fixed. 
%In several security proofs with finite-key analysis, 
%the mehod with simple random sampling is used assuming the protcol where the predetermined size of bit strings 
%are obtained after sifting process. 
In several finite-key analyses~\cite{2014Hayashi,2015Lucamarini} based on simple random sampling, 
efforts were made to find bounds on hypergeometric distribution 
which are related to binomial distribution in order to 
simplify numerical calculation.

In order to mitigate the inefficiency arising from basis mismatch between the sender and the receiver, 
the BB84 protocol is often implemented with biased basis choice~\cite{2004Lo}, 
in which the minor basis is used solely for monitoring leaked information in the major basis. 
In this case, the whole data from the rounds in the monitoring basis is regarded as a sample, 
with each round selected with a constant probability dictated in the protocol as that of the basis choice. 
This suggests that 
the data from the monitoring basis is related to Bernoulli sampling, 
in which each element of the population of size $n_2$ is sampled with fixed probability $p_1$. 
The number of samples $n_1$ obeys binomial distribution 
\begin{equation} 
{\rm BI}(n_1;n_2,p_1):=\binom{n_2}{n_1}p_1^{n_1}(1-p_1)^{n_2-n_1}. 
\end{equation}
If the BB84 protocol with biased basis choice essentially includes the property of 
the binomial distribution, 
analysis based on the conventional simple random sampling may introduce unnecessary complexity 
and possibly leads to a lower key rate.

In this paper, we work on the finite-key analysis by focusing on the Bernoulli sampling 
instead of simple random sampling. 
We propose the method based on binomial distribution which is 
parametrized by the basis choice probabilities in the protocol. 
Differently from the previous works which deal with binomial distribution to derive bounds on 
hypergeometric distribution, 
our work is based on binomial distributions inherent in the protocol. 
Our method is especially suited for the BB84 protocol with weak coherent pulses (WCP), 
providing a simpler analysis with less 
estimation processes as well as achieving a higher key rate compared to the analysis 
with simple random sampling. 
We further apply this method to the differential quadrature phase shift (DQPS) protocol \cite{2009Inoue}, whose 
security was proved recently in the asymptotic regime \cite{2016Kawakami}. 
As a result, we show that the advantage of the DQPS protocol over the phase-encoded BB84 protocol with WCP 
still remains in the finite key regime. 

This paper is organized as follows. In Sec.~\ref{protocol}, we describe details of the 
BB84 protocol which is considered in this work, along with a summary of notations used in this paper. 
In Sec.~\ref{idealBB84}, we propose a method of finite-key analysis based on Bernoulli sampling, 
and applies it to the ideal BB84 protocol 
where Alice and Bob can manipulate perfect single photon states. 
The proposed method is then applied to the BB84 protocol with WCP as well as the DQPS protocol 
in Sec.~\ref{WCPBB84}. 
Finally, we give discussion and conclusion in Sec.~\ref{conclusion}.

\section{BB84 Protocol}\label{protocol}
In most part of this paper, we discuss the finite key analysis of the BB84 protocol~\cite{Bennett1984}, 
which is given in the following. 
The sender Alice and the receiver Bob independently chooses two bases ($Z$ basis and $X$ basis) 
with a biased probability. 
The final key is generated only from $Z$-basis data, while $X$-basis data is used for 
leak monitoring to determine the amount for privacy amplification. 
The number of total rounds $n_{\rm rep}$ are predetermined and 
there is no threshold for data size after sifting process, 
which means that the size of sifted key $n_Z$ and that of monitoring bits $n_X$ are not 
determined until quantum communication is over. 
%We assume that the error correction is done by exchanging $\lambda_{\rm EC}$ bits 
%(e.g. syndrome of Alice's sifted key) encrypted by consuming the same length of pre-obtained secret key.  

The protocol proceeds as follows with predetermined parameters $\tilde{p}_Z$, $\tilde{p}_X=1-\tilde{p}_Z$ 
and $n_{\rm rep}$. In its description, $|\bm{\kappa}|$ represents the length of a bit sequence 
$\bm{\kappa}$.\\\\
(1) Alice chooses $Z$ basis or $X$ basis with probability $\tilde{p}_Z$ and 
$\tilde{p}_X$, 
respectively. 
She chooses a uniformly random bit $\{0,1\}$.\\
(2) Alice prepares one of states $\{\hat{\rho}_{Z,0},\hat{\rho}_{Z,1},\hat{\rho}_{X,0},\hat{\rho}_{X,1}\}$ based on 
the selected basis and bit. 
 She sends the prepared state to Bob over the quantum channel. 
\\
(3)  Bob chooses $Z$ basis or $X$ basis 
with probability $\tilde{p}_Z$ and 
$\tilde{p}_X$, 
respectively. 
He measures a received state in chosen basis and obtains the outcome \{0, 1,  no-detection\}. \\	
(4) They repeat the sequence (1) to (3), which we call a round, by $n_{\rm rep}$ times.\\
(5) Bob publicly announces whether each round has resulted in a detection or not. 
Let $n_{\rm det}$ be the number of rounds with detection. \\
(6) 
Alice and Bob disclose all of their basis choices. 
Among the $n_{\rm det}$ detected rounds,  
the rounds where both Alice and Bob chose the $Z$ basis are called ``$Z$-labeled'' rounds, and 
the rounds where they chose the $X$ basis are called ``$X$-labeled'' rounds. 
They define sifted keys $\bm{\kappa}_{A,Z}$ and $\bm{\kappa}_{B,Z}$ 
by concatenating the bits for the $Z$-labeled rounds, and similarly define 
$\bm{\kappa}_{A,X}$ and $\bm{\kappa}_{B,X}$ for the $X$-labeled rounds. 
Let their sizes be $n_Z:=|\bm{\kappa}_{A,Z}|=|\bm{\kappa}_{B,Z}|$ and 
$n_X:=|\bm{\kappa}_{A,X}|=|\bm{\kappa}_{B,X}|$. 
\\
(7) 
They disclose and compare $\bm{\kappa}_{A,X}$ and $\bm{\kappa}_{B,X}$ to determine the number of bit errors $k_X$ 
included in them.\\ 
(8) Through public discussion, 
Bob corrects his keys $\bm{\kappa}_{B, Z}$ to make it coincide 
with Alice's key $\bm{\kappa}_{A, Z}$ and obtains $\bm{\kappa}_{B, Z}^{\rm cor}$ 
$(|\bm{\kappa}_{B, Z}^{\rm cor}|=n_Z)$. \\
(9) Alice and Bob conduct privacy amplification by shortening $\bm{\kappa}_{A,Z}$ and 
$\bm{\kappa}_{B,Z}^{\rm cor}$ to obtain final keys $\bm{\kappa}_{A,Z}^{\rm fin}$ and 
$\bm{\kappa}_{B,Z}^{\rm fin}$ 
of size $n_{\rm fin}$. \\

In the subsequent sections, we discuss how we can determine the final key length $n_{\rm fin}$ 
as a function of the random variables $n_Z$, $n_X$, and $k_X$ obtained in the protocol to satisfy a given 
security criteria. 
For convenience, we define several variables and parameters as 
$n_{\rm tot}:=n_X+n_Z$ and 
\begin{align}
p_Z:=&\tilde{p}_Z^2/(\tilde{p}_Z^2+\tilde{p}_X^2),\nonumber 
\\ 
p_X:=&\tilde{p}_X^2/(\tilde{p}_Z^2+\tilde{p}_X^2).
\label{pdef}
\end{align} 
Throughout this paper, we adopt an abuse of notation to use the same symbol for a random variable 
$\tilde{n}$ and its value $n$, whenever the distinction is obvious. 
For example, we denote ${\rm Pr}(n>3)$ instead of ${\rm Pr}(\tilde{n}>3)$. 
We denote by ${\rm Pr}(n)$ the probability mass function 
${\rm Pr}(\tilde{n}=n)$. 
Similarly, we use ${\rm Pr}(n \mid m)$ instead of  ${\rm Pr}(\tilde{n}=n \mid \tilde{m}=m)$. 
We define $\ket{0_Z}$ and $\ket{1_Z}$ as basis vectors of $Z$ basis on a qubit system, 
and $\ket{0_X}:=(\ket{0_Z}+\ket{1_Z})/\sqrt{2}$ and $\ket{1_X}:=(\ket{0_Z}-\ket{1_Z})/\sqrt{2}$ 
as those of $X$ basis. 
When the same notations are used for an optical signal, 
it should be understood that they refer to the states in the subspace of a single photon contained 
in two modes, such as polarizations. 
The four Bell sates are represented as 
$\ket{\Phi^{\pm}}$ and $\ket{\Psi^{\pm}}$ where 
\begin{align}
\ket{\Phi^\pm}&:=\frac{1}{\sqrt{2}}(\ket{00_Z}\pm \ket{11_Z}),\\
\ket{\Psi^\pm}&:=\frac{1}{\sqrt{2}}(\ket{01_Z}\pm \ket{10_Z}). 
\end{align}
We define a function $h(x)$ for $x\geq 0$ as 
\begin{eqnarray}
h(x)=
\left \{
\begin{matrix}
-x {\rm log}_2 x-(1-x) {\rm log}_2(1-x) & (0\leq x \leq 1/2)\\
1 & (x>1/2).\\
\end{matrix}\right.
\end{eqnarray}

%We consider the security of the above protocol under the following assumptions. 
%At Alice's site, we assume  $\rho_{Z,0}+\rho_{Z,1}=\rho_{X,0}+\rho_{X,1}$. 
%At Bob's site, we assume that 
%the probability that a signal is detected at Bob's receiver is independent of 
%his basis choice. 
%These assumptions mean that Eve can not extract the information of basis that Alice and Bob selected. 

%In the above protocol, the length of the final key depends on variable number 
% data $n_Z$ and sample bits $n_X$ , 
%which differentiates the protocol from the usual one in which the 
%length of final key depends on fixed data-size $\bar{n}_Z$ and $\bar{n}_X$. 
%In this paper, we call the former porotocol ``variable key-length protocol'', and call the latter 
%``fixed key-length protocol''.
%In the fixed key-length protocol, the protocol lasts until both 
%$n_Z\geq \bar{n}_Z$ and $n_X\geq \bar{n}_X$ are satisfied, 
%followed by randomly choosing 
%$\bar{n}_Z$ bits from $n_Z$ bits of data basis or 
%$\bar{n}_Z$ bits from $n_X$ bits of check basis. 

\section{Analysis for the ideal BB84 protocol}\label{idealBB84}
Here we consider finite-key analysis for the ideal qubit-based  
 BB84 protocol, in which Alice sends 
 a single photon in the states 
 $\{\hat{\rho}_{W,a}=\ket{a_W}\bra{a_W}_S\}$ $(W\in \{Z,X\}, a\in\{0,1\})$
 in Step~(2) and Bob conducts ideal measurement with unit efficiency 
 described by POVM (positive operator-valued measure) 
 $\{\ket{0_W}\bra{0_W}_S,\ket{1_W}\bra{1_W}_S,\hat{\mathbbm{1}}_S-\ket{0_W}\bra{0_W}_S-\ket{1_W}\bra{1_W}_S\}$ 
 corresponding to the outcome \{0,1,{\rm no-detection}\}.

\subsection{Security criteria and formalism for key length}\label{formalism}
In this work, we follow the security definition based on universally composable security 
\cite{2009Benor,2009Renner}. 
The protocol is called $\epsilon_{\rm sec}$ secure if it is both $\epsilon_{\rm c}$-correct and 
$\epsilon_{\rm s}$-secret where $\epsilon_{\rm sec}=\epsilon_{\rm c}+\epsilon_{\rm s}$. 
We call the protocol is $\epsilon_{\rm c}$-correct if 
${\rm Pr}(\bm{\kappa}_{A,Z}^{\rm fin}\neq \bm{\kappa}_{B,Z}^{\rm fin})\leq \epsilon_{\rm c}$ holds.
We call the protocol is $\epsilon_{\rm s}$-secret if  
\begin{align}
\frac{1}{2}\big|\big|~\hat{\rho}^{\rm fin}_{AE}-\hat{\rho}^{\rm ideal}_{AE}~\big|\big|\leq \epsilon_{\rm s} 
\label{tracedistance}
\end{align}
where 
$\hat{\rho}^{\rm fin}_{AE}:=\sum_{\bm{\kappa}_{A,Z}^{\rm fin}}p(\bm{\kappa}_{A,Z}^{\rm fin})
\ket{\bm{\kappa}_{A,Z}^{\rm fin}}\bra{\bm{\kappa}_{A,Z}^{\rm fin}}\otimes \hat{\rho}_E(\bm{\kappa}_{A,Z}^{\rm fin})$
is a classical-quantum state between Alice's key and Eve's system  after finishing the protocol 
and 
$\hat{\rho}^{\rm ideal}_{AE}$ 
is an ideal separable state in which Alice's key is uniformly distributed over
$2^{|\bm{\kappa}_{A,Z}^{\rm fin}|}$ values and decoupled from Eve's system.

%$\hat{\rho}^{(\rm ideal)}_{AE}:=\sum_{\bm{\kappa}_{A,Z}^{(\rm fin)}}\frac{1}{2^{|\bm{\kappa}_{A,Z}^{(\rm fin)}|}}\ket{\bm{\kappa}_{A,Z}^{(\rm fin)}}\bra{\bm{\kappa}_{A,Z}^{(\rm fin)}}\otimes \hat{\rho}_E$ 
%is an ideal separable state in which no information of Alice's final keys is leaked to Eve. 

In the ideal BB84 protocol, 
Alice's procedure of selecting a random bit and a basis, and preparing the corresponding signal can be 
replaced~\cite{BBM92} 
by preparation of $\ket{\Phi^+}_{AS}$ followed by the measurement on the system $A$ 
on $\{\ket{0_Z}_A,\ket{1_Z}_A\}$ ($Z$-basis) or on $\{\ket{0_X}_A,\ket{1_X}_A\}$ ($X$-basis). 
Bob's measurement is also replaced by 
a filtering operation to make sure a single photon is received and transfer its state to a qubit $B$, 
followed by the orthogonal measurement of $B$ on the chosen basis to determine the outcome 0 or 1. 
When the filtering fails, the outcome is ``no-detection''. 
According to Ref.~\cite{1999Lo,2000Shor}, 
a phase error occurs when Alice and Bob conduct virtual Bell-basis measurement 
on a $Z$-labeled round after Eve's intervention 
to obtain the outcome for  
$\ket{\Phi^-}\bra{\Phi^-}_{AB}$ or $\ket{\Psi^-}\bra{\Psi^-}_{AB}$. 
Since we have the relation 
\begin{equation}
\ket{\Phi^-}\bra{\Phi^-}_{AB}+\ket{\Psi^-}\bra{\Psi^-}_{AB}=\ket{01_X}\bra{01_X}_{AB}+\ket{10_X}\bra{10_X}_{AB}, 
\end{equation}
phase error is equivalently defined as a bit error 
which occurs when Alice and Bob conduct virtual $X$-basis measurement after Eve's intervention on a 
$Z$-labeled round. 
An important property which will be used in the next subsection is that the measurement for a phase error 
on a $Z$-labeled round and the measurement for a bit error on an $X$-labeled round are identical, 
and hence they only differs in the labeling.

Let $k_{\rm ph}$ be a random variable which represents the number of phase errors on 
$n_Z$ $Z$-labeled rounds. 
Once we have a good upper bound on $k_{\rm ph}$, a secure key length can be calculated as follows.
Suppose that we have a function $f(k_X,n_X,n_{\rm tot})$ which satisfies 
\begin{align}
{{\rm Pr}(k_{\rm ph}> f(k_X,n_X,n_{\rm tot}))\leq \epsilon_{\rm PE}} \label{f} 
 \end{align} 
regardless of Eve's attack strategy. 
By setting \cite{2014Hayashi} 
\begin{equation}
\epsilon_{\rm s}=\sqrt{2}\sqrt{\epsilon_{\rm PE}+\epsilon_{\rm PA}}, 
\label{epsps}
\end{equation}
the protocol is $\epsilon_{\rm c}$-correct and $\epsilon_{\rm s}$-secret  
if the final key length $n_{\rm fin}$ satisfies 
\begin{equation} 
n_{\rm fin}\leq  
n_Z - 
\left \lceil
n_Z h\left (\frac{f(k_X,n_X,n_{\rm tot})}{n_Z}\right) 
+ {\rm log_2} \frac{1}{\epsilon_{\rm PA} }
\right \rceil
-\lambda_{\rm EC}(\epsilon_{\rm c}).
\end{equation}
where $\lceil~\rceil$ represents the ceiling function and 
$\lambda_{\rm EC}(\epsilon_{\rm c})$ is the cost of error correction to 
achieve $\epsilon_{\rm c}$-correctness.  
For simplicity, we will replace the right-hand side by a slightly pessimistic bound as 
\begin{equation} 
n_{\rm fin}\leq  n_Z(1-h\left (\frac{f(k_X,n_X,n_{\rm tot})}{n_Z}\right))
-{\rm log_2} \frac{2}{\epsilon_{\rm PA}} - \lambda_{\rm EC}(\epsilon_{\rm c}). 
\label{keyrate}
\end{equation}

%\be
%\left \lceil 
%{\rm log_2} \frac{1}{\epsilon_1}
%\right \rceil
%+
%\left \lceil 
%{\rm log_2} \frac{1}{\epsilon_2}
%\right \rceil
%\leq 
%\left \lceil 
%{\rm log_2} \frac{2}{\epsilon_1 \epsilon_2}
%\right \rceil.
%\ee

\subsection{Bounds on phase errors}\label{bounds} 
In this subsection, we discuss the specific methods to obtain $f(k_X,n_X,n_{\rm tot})$ in Eq.~(\ref{f}) 
including a method based on the Bernoulli sampling, and a more conventional method based on the 
simple random sampling. 
We also introduce a third, rather convoluted 
method, which will help to elucidate the difference between the former two methods.

Before discussing each of the methods, we first derive general statistical properties. 
Since the $Z$-labeled phase error and the $X$-labeled bit error are obtained by identical measurements, 
the procedure to obtain those errors is equivalent to the following steps 
after discarding the rounds with no-detection (i.e., with Bob failing to receive a qubit): 
(a) Alice and Bob further discard each of the remaining rounds  
with probability $1-\tilde{p}_Z^2-\tilde{p}_X^2$. 
(b) They make $X$-basis measurements on the remaining $n_{\rm tot}$ rounds 
and obtain $k_{\rm tot}$ errors. 
(c) Finally, they label each of the $n_{\rm tot}$ rounds as $Z$ or $X$ with probability 
$p_Z$ and $p_X$ (see \eq{pdef}),  
respectively, and obtain $k_{\rm ph}$ phase errors in $Z$-labeled rounds and $k_X=k_{\rm tot}-k_{\rm ph}$ 
bit errors 
in $X$-labeled rounds. 
In this procedure, 
since $k_X$ errors are sampled from $k_{\rm tot}$ errors with a fixed probability $p_X$, 
it follows a binomial distribution if $k_{\rm tot}$ and $n_{\rm tot}$ are fixed: 
\begin{equation}
{\rm Pr}(k_X \mid k_{\rm tot},n_{\rm tot})={\rm BI}(k_X;k_{\rm tot},p_X).\label{BIk}
\end{equation} 
On the other hand, the step (c) of the above procedure is equivalently denoted as follows: 
Alice and Bob draw a number $n_X$ based on the binomial distribution 
${\rm BI}(n_X;n_{\rm tot},p_X)$, 
and then select $n_X$ random rounds among the $n_{\rm tot}$ rounds to label as $X$, thereby determining $k_X$. 
This implies that the number $k_X$ obeys hypergeometric distribution if 
$n_X$, $k_{\rm tot}$ and $n_{\rm tot}$ are fixed: 
 \begin{align}
 {\rm Pr}(k_X \mid n_X,k_{\rm tot},n_{\rm tot})={\rm HG}(k_X;n_X,k_{\rm tot},n_{\rm tot}).\label{HG}
 \end{align}

In order to use the properties derived above, it is convenient to reformulate \eq{f} as follows. 
From Eq.~(\ref{f}), we have 
\begin{equation}
\sum_{k_{\rm tot},n_{\rm tot}}
{\rm Pr}(k_{\rm ph} > f(k_X,n_X,n_{\rm tot}) \mid k_{\rm tot},n_{\rm tot})
{\rm Pr}(k_{\rm tot},n_{\rm tot})
\leq \epsilon_{\rm PE}.
\end{equation} 
Since ${\rm Pr}(k_{\rm tot},n_{\rm tot})$ can be under control of Eve, we seek 
 for $f(k_X,n_X,n_{\rm tot})$ satisfying 
\begin{equation}
{{\rm Pr}(k_{\rm ph} > f(k_X,n_X,n_{\rm tot}) \mid k_{\rm tot},n_{\rm tot})\leq \epsilon_{\rm PE}} 
\label{final}
 \end{equation} 
 for any $k_{\rm tot}$ and $n_{\rm tot}$, 
 which is a sufficient condition for Eq.~(\ref{f}). 
For later convenience, we equivalently describe Eq.~(\ref{final}) as 
\begin{equation}
\sum_{k_X,n_X;k_X < k_{\rm tot}-f(k_X,n_X,n_{\rm tot})}
{\rm Pr}(k_X,n_X \mid k_{\rm tot},n_{\rm tot}) 
\leq \epsilon_{\rm PE}. 
\label{ff}
\end{equation}

The first method to determine $f(k_X,n_X,n_{\rm tot})$, whose utility we will emphasize throughout this paper, 
is based on Bernoulli sampling
using the property 
of binomial distribution \eq{BIk}. 
This method adopts $f(k_X,n_X,n_{\rm tot})=f_{\rm BI}(k_X)$ where 
\begin{align}
&f_{\rm BI}(k_X)
:={\rm min}
\Bigg\{ k_{\rm tot}\Big| 
C_{\rm BI}(k_X;k_{\rm tot},p_X)
 \leq \epsilon_{\rm PE} \Bigg \}-k_X-1
\label{fb} \\
&C_{\rm BI}(k_X;k_{\rm tot},p_X):=\sum_{k_X'\leq k_X}{\rm BI}(k_X';k_{\rm tot},p_X)\label{cb}. 
\end{align}
The proof that $f_{\rm BI}(k_X)$ 
 satisfies Eq.~(\ref{final}) is as follows. 
Let 
$\mbar{k
}_X(k_{\rm tot}):={\rm max}\{k_X \mid k_{\rm tot}> f_{\rm BI}(k_X)+k_X\}$. 
Then we have 
\begin{equation}
\sum_{k_X;~k_{\rm tot}> f_{\rm BI}(k_X)+k_X} {\rm BI}(k_X;k_{\rm tot},p_X) 
\leq 
C_{\rm BI}(\mbar{k}_X(k_{\rm tot});k_{\rm tot},p_X)
\label{tahoo}.
\end{equation} 
Since $C_{\rm BI}(k_X;k_{\rm tot},p_X)$ is 
a decreasing function of $k_{\rm tot}$, from Eq.~(\ref{fb}) we have 
$C_{\rm BI}(k_X;k_{\rm tot},p_X) \leq \epsilon_{\rm PE}$ for any pair $(k_X,k_{\rm tot})$ 
satisfying $k_{\rm tot}\geq f_{\rm BI}(k_X)+k_X+1$. 
Since $k_{\rm tot}\geq f_{\rm BI}(\mbar{k}_X(k_{\rm tot}))+\mbar{k}_X(k_{\rm tot})+1$ 
holds by definition of $\mbar{k}_X(k_{\rm tot})$, 
we have $C_{\rm BI}(\mbar{k}_X(k_{\rm tot});k_{\rm tot},p_X) \leq \epsilon_{\rm PE}$. 
By connecting this to Eq.~(\ref{tahoo}), we have 
\begin{align}
\sum_{k_X;~k_X < k_{\rm tot}- f_{\rm BI}(k_X)} {\rm BI}(k_X;k_{\rm tot},p_X)
\leq 
\epsilon_{\rm PE},
\label{opa}
\end{align}
for any $k_{\rm tot}$. 
From Eqs.~(\ref{BIk}) and (\ref{opa}), we have 
\begin{align}
&\sum_{k_X,n_X;~ k_X < k_{\rm tot}-f_{\rm BI}(k_X)}{\rm Pr}(k_X,n_X \mid k_{\rm tot},n_{\rm tot}) \nonumber\\
&=\sum_{k_X;~k_X < k_{\rm tot}-f_{\rm BI}(k_X)} {\rm Pr}(k_X \mid k_{\rm tot},n_{\rm tot})\nonumber \\
&\leq \epsilon_{\rm PE},
\label{ffbi}
\end{align}
which 
is identical to Eq.~(\ref{ff}) with $f(k_X,n_X,n_{\rm tot})=f_{\rm BI}(k_X)$. 
Therefore, we have 
\begin{equation}
{{\rm Pr}(k_{\rm ph} > f_{\rm BI}(k_X) \mid k_{\rm tot},n_{\rm tot})\leq \epsilon_{\rm PE}}. 
\label{ffbi}
 \end{equation} 
As a result of the Bernoulli-sampling method,  
the protocol is $\epsilon_{\rm c}$-correct and $\epsilon_{\rm s}$-secret 
if the final key length $n_{\rm fin}$ satisfies 
\begin{equation} 
n_{\rm fin} \leq l^{(\rm BI)}:= n_Z(1-h\left (\frac{f_{\rm BI}(k_X)}{n_Z}\right))
-{\rm log_2} \frac{2}{\epsilon_{\rm PA}} - \lambda_{\rm EC}(\epsilon_{\rm c}).
\label{BIkeyrate}
\end{equation}
where $\epsilon_{\rm s}$ is given by Eq.~(\ref{epsps}).

The second method is based on simple random sampling, applying  
the property of the hypergeometric distribution \eq{HG}, 
which is already seen in Ref.~\cite{2012Tomamichel,2014Wen,2014Curty,2014Hayashi}, for example. 
This method adopts 
 $f(k_X,n_X,n_{\rm tot})=f_{\rm HG}(k_X,n_X,n_{\rm tot})$  where
\begin{align}
f_{{\rm HG}}(k_X,n_X,n_{\rm tot}):=
{\rm min}&
\Bigg\{ k_{\rm tot}\Big| C_{\rm HG}(k_X;n_X,k_{\rm tot},n_{\rm tot}) \leq \epsilon_{\rm PE} \Bigg \}
\nonumber\\
&~~~~~~~~~~~~~~~~~~~~~~~~~~~~~~~~~~~~~~~~~~-k_X -1
\nonumber \\
C_{\rm HG}(k_X;n_X,k_{\rm tot},n_{\rm tot}):=&\sum_{k_X'\leq k_X}{\rm HG}(k_X';n_X, k_{\rm tot},n_{\rm tot}). 
\label{fhg}
\end{align}
%This holds from the following statements. 
%First, the expected value of $k_X$ is monotonically increasing as $k_{\rm tot}$ gets large 
%regardless of the values $n_{\rm tot}$ and $n_X$. 
%Thus $S_{{\rm HG}}(k_{\rm tot},n_X,n_{\rm tot})$ is a monotonically decreasing function of $k_{\rm tot}$, which 
%means $k_X\leq S_{{\rm HG}}(k_{\rm tot},n_X,n_{\rm tot}) \leftrightarrow
%k_{\rm tot}\leq S_{{\rm HG}}^{-1}(k_X,n_X,n_{\rm tot})$. 
%Therefore, from 
%\begin{align}
%\sum_{\{k_X| k_X\leq S_{{\rm HG}}(k_{\rm tot},n_X,n_{\rm tot})\}} 
%{\rm HG}(k_X;k_{\rm tot},n_X,n_{\rm tot}) 
%\leq \epsilon_{\rm PE} 
%\end{align}
%we obtain 
%\begin{align}
%\sum_{\{k_X| S_{{\rm HG}}^{-1}(k_X,n_X,n_{\rm tot})\leq 
%k_{\rm tot}\}} {\rm HG}(k_X;k_{\rm tot},n_X,n_{\rm tot}) 
%\leq \epsilon_{\rm PE}.
%\end{align}
%From the definition of $k_{\rm ph}$ and $f_{{\rm HG}}(k_X,n_X,n_{\rm tot})$, 
%\begin{align}
%\sum_{\{k_{\rm ph} \geq f_{{\rm HG}}(k_X,n_X,n_{\rm tot})\}}{\rm HG}(k_X;k_{\rm tot},n_X,n_{\rm tot})\leq \epsilon_{\rm PE},
%\end{align}
%which leads to Eq.~(\ref{ff}). 
The proof that $f_{{\rm HG}}(k_X,n_X,n_{\rm tot})$ 
 satisfies Eq.~(\ref{final}) is similar to the proof for $f_{\rm BI}(k_X)$. 
Recall that the proof for $f_{\rm BI}(k_X)$ did not use the explicit form of 
${\rm BI}(k_X',k_{\rm tot},p_X)$ 
but only used the decreasing property of $C_{\rm BI}(k_X;k_{\rm tot},p_X)$ as a function of 
$k_{\rm tot}$. 
Since $C_{\rm HG}(k_X;n_X,k_{\rm tot},n_{\rm tot})$ 
is also a decreasing function of $k_{\rm tot}$, 
we have 
\begin{align}
\sum_{k_X;~k_X < k_{\rm tot}- f_{\rm HG}(k_X,n_X,n_{\rm tot})} {\rm HG}(k_X;n_X,k_{\rm tot},n_{\rm tot})
\leq 
\epsilon_{\rm PE}
\label{opaopa}
\end{align}
for any $n_X$, $k_{\rm tot}$ and $n_{\rm tot}$, which is analogous to Eq.~(\ref{opa}). 
From Eqs.~(\ref{HG}) and (\ref{opaopa}), we have 
\begin{align}
&\sum_{k_X,n_X;~k_X < k_{\rm tot}-f_{\rm HG}(k_X,n_X,n_{\rm tot})}{\rm Pr}(k_X,n_X \mid k_{\rm tot},n_{\rm tot})
\nonumber \\
&=\sum_{k_X,n_X;~k_X < k_{\rm tot}-f_{\rm HG}(k_X,n_X,n_{\rm tot})}{\rm Pr}(k_X \mid n_X,k_{\rm tot},n_{\rm tot})
{\rm Pr}(n_X \mid k_{\rm tot},n_{\rm tot})
\nonumber \\
&\leq \epsilon_{\rm PE},
\end{align}
which is identical to Eq.~(\ref{ff}) with $f(k_X,n_X,n_{\rm tot})=f_{\rm HG}(k_X,n_X,n_{\rm tot})$. 
Therefore, we have 
\begin{equation}
{{\rm Pr}(k_{\rm ph} > f_{\rm HG}(k_X,n_X,n_{\rm tot}) \mid k_{\rm tot},n_{\rm tot})\leq \epsilon_{\rm PE}}. 
\label{ffhg}
 \end{equation} 
As a result of the method with simple random sampling,  
the protocol is $\epsilon_{\rm c}$-correct and $\epsilon_{\rm s}$-secret 
if the secret key length $n_{\rm fin}$ satisfies 
\begin{equation} 
n_{\rm fin} \leq l^{(\rm HG)}:= n_Z(1-h\left (\frac{f_{\rm HG}(k_X,n_X,n_{\rm tot})}{n_Z}\right))
-{\rm log_2} \frac{2}{\epsilon_{\rm PA}} - \lambda_{\rm EC}(\epsilon_{\rm c})
\label{HGkeyrate}
\end{equation}
where $\epsilon_{\rm s}$ is given by Eq.~(\ref{epsps}).

To understand the relation between the two methods with 
Bernoulli sampling and simple random sampling, 
we introduce another 
method which uses full knowledge of the distribution 
${\rm Pr}(k_X,n_X \mid k_{\rm tot},n_{\rm tot})$ 
appearing in \eq{ff}.  
The argument before \eq{BIk} also implies that the number 
$m_X:=n_X-k_X$ of $X$-labeled rounds without bit error obeys binomial distribution 
${\rm BI}(m_X;n_{\rm tot}-k_{\rm tot},p_X)$, 
and that $m_X$ and $k_X$ are independent conditioned on
$k_{\rm tot}$ and $n_{\rm tot}$. 
We thus obtain 
\begin{align}
{\rm Pr}(k_X,n_X \mid k_{\rm tot},n_{\rm tot})
={\rm BI}(k_X;k_{\rm tot},p_X){\rm BI}(n_X-k_X;n_{\rm tot}-k_{\rm tot},p_X)
\label{kebab2}. 
\end{align}
The argument leading to \eq{HG} gives another expression for the distribution as 
\begin{align}
{\rm Pr}(k_X,n_X \mid k_{\rm tot},n_{\rm tot})
=
{\rm HG}(k_X;n_X,k_{\rm tot},n_{\rm tot})
{\rm BI}(n_X;n_{\rm tot},p_X).
\end{align}
As a result, Eq.~(\ref{ff}) is expressed in the following two equivalent ways:  
\begin{align}
&\sum_{k_X,m_X;k_X < k_{\rm tot}-f(k_X,k_X+m_X,n_{\rm tot})}
{\rm BI}(k_X;k_{\rm tot},p_X){\rm BI}(m_X;n_{\rm tot}-k_{\rm tot},p_X)
\nonumber \\
&~~~~~~~~~~~~~~~~~~~~~~~~~~~~~~~~~~~~~~~~~~~~~~~~~~~~~~~~~~~~~~~~~~~~~~~~~~~~~~~~~~~~\leq \epsilon_{\rm PE}. 
\nonumber \\
\label{ikko}
\end{align}
or
\begin{align}
\sum_{k_X,n_X;k_X < k_{\rm tot}-f(k_X,n_X,n_{\rm tot})}
{\rm HG}(k_X;n_X,k_{\rm tot},n_{\rm tot}){\rm BI}(n_X;n_{\rm tot},p_X) 
\leq \epsilon_{\rm PE}.
\label{niko}
\end{align}

Since $f_{\rm BI}(k_X)$ satisfies Eq.~(\ref{opa}), 
 Eq.~(\ref{ikko}) holds if $f(k_X,k_X+m_X,n_{\rm tot})=f_{\rm BI}(k_X)$. 
Similarly, 
since $f_{\rm HG}(k_X,n_X,n_{\rm tot})$ satisfies Eq.~(\ref{opaopa}), 
 Eq.~(\ref{niko}) holds 
if $f(k_X,n_X,n_{\rm tot})=f_{\rm HG}(k_X,n_X,n_{\rm tot})$. 
On the other hand, the condition of Eqs.~(\ref{ikko}) and (\ref{niko}) do not imply Eq.~(\ref{opa}) 
or Eq.~(\ref{opaopa}). 
Therefore, there could be a better bound compared to 
$f_{\rm BI}(k_X)$ and $f_{\rm HG}(k_X,n_X,n_{\rm tot})$ based on 
Eq.~(\ref{ikko}) or Eq.~(\ref{niko}). 
In general, it is very complicated to determine the optimal function 
$f(k_X,n_X,n_{\rm tot})$ for the final key length $n_{\rm fin}$, 
since it will depend on the explicit functional dependence of $n_{\rm fin}$ on $f(k_X,n_X,n_{\rm tot})$.

The difference between the two equivalent conditions Eqs.~(\ref{ikko}) and (\ref{niko}) is 
the choice of two variables from three no-independent random variables $k_X$, $n_X$ and $m_X$. 
When $(k_X,n_X)$ are chosen in Eq.~(\ref{niko}), the distribution of $k_X$, 
${\rm HG}(k_X;n_X,k_{\rm tot},n_{\rm tot})$ is dependent on the value of 
$n_X$. 
On the other hand, Eq.~(\ref{ikko}) implies that two variables 
$(k_X,m_X)$ are independent of each other. 
This suggests that the underlying statistics in the 
BB84 protocol with biased basis choice are understood in terms of independent binomial distributions.

Let us mention the difference from the other works~\cite{2014Hayashi,2015Lucamarini} 
which deal with relations between bounds on binomial distribution and ones on hypergeometric distribution 
since the former are easily treated with existing mathematical packages. 
%The binomial distributions appearing in those works are 
%${\rm BI}(k_X;n_X,k_{\rm tot}/n_{\rm tot})$ and ${\rm BI}(k_X;k_{\rm tot},n_X/n_{\rm tot})$ in our notation. 
Ref.~\cite{2014Hayashi} uses the property, which dates back to Hoeffding~\cite{1963Hoeffding}, 
that expectation of a convex function over hypergeometric dirstribution is no larger than that over 
binomial distribution. 
In \cite{2015Lucamarini}, Ahrens map~\cite{1987Ahrens} was used to show that hypergeometric dirstribution is bounded by a permutated 
binomial distribution  within a factor of $\sqrt{2}$. 
In contrast to these works, in our case 
%The former is the distribution of the number of 
%errors $k_X$ in $n_X$ samples 
%drawn by the simple random sampling {\it with replacement}. 
%Hence its cumulative function $C_{\rm BI}(k_X;n_X,k_{\rm tot}/n_{\rm tot})$ should thus be no smaller than that 
%of hypergeometric distribution ${\rm HG}(k_X;n_X,k_{\rm tot},n_{\rm tot})$ 
%(obtained by the simple random sampling $without$ $replacement$), leading to a  
%pessimistic bound of secure key rate. 
%Similarly, $C_{\rm BI}(k_X;k_{\rm tot},n_X/n_{\rm tot})$ is also an upper bound of 
%$C_{\rm HG}(k_X;n_X,k_{\rm tot},n_{\rm tot})$ since 
%${\rm HG}(k_X;n_X,k_{\rm tot},n_{\rm tot})={\rm HG}(k_X;k_{\rm tot},n_X,n_{\rm tot})$.  
%Hence the previous works have introduced a binomial distributions as a bound on the true distribution, 
%which is in stark contrast to the fact that
the probability distribution \eq{BIk} 
reflects the binomial distribution inherent in the BB84 protocol with biased basis choice.

\subsection{Numerical examples}
Here we numerically compare the final key lengths derived from the three methods in the 
last subsection in the simplest cases. 
We calculate the key lengths for the case where 
no error is observed ($k_X=0$) 
and every signal is detected ($n_{\rm tot}=n_{\rm rep})$. 
The cost of error correction is set to 
$\lambda_{\rm EC}(\epsilon_{\rm c})={\rm log}_2 (1/\epsilon_{\rm c})$. 
We also assume $n_Z=n_{\rm rep}\tilde{p}_Z^2$ and $n_X=n_{\rm rep}\tilde{p}_X^2$. 

If we do not care about the key length for $k_X>0$, the optimal choice of $f(k_X,n_X,n_{\rm tot})$ 
satisfying Eq.~(\ref{niko}) (or Eq.~(\ref{ikko})) is given by 
$f(k_X,n_X,n_{\rm tot})=n_{\rm tot}-n_X$ for $k_X \geq 1$ and 
$f(0,n_X,n_{\rm tot})=f_{\rm opt}^{(k_X=0)}(n_X,n_{\rm tot})$
with 
\begin{align}
&f_{\rm opt}^{(k_X=0)}(n_X,n_{\rm tot}):=
{\rm min}\Bigg \{k_{\rm tot} \Big| 
G(n_X;k_{\rm tot},n_{\rm tot})
\leq \epsilon_{\rm PE} \Bigg \}-1 \nonumber \\
&G(n_X;k_{\rm tot},n_{\rm tot}) \nonumber \\
&~~~~:=
\sum_{n_X\leq n_X'\leq n_{\rm tot}-k_{\rm tot}}
{\rm HG}(0;n_X',k_{\rm tot},n_{\rm tot}){\rm BI}(n_X';n_{\rm tot},p_X)
.
\label{ftight}
\end{align}
The proof is 
analogous to the one for $f_{\rm BI}(k_X)$ or $f_{\rm HG}(k_X,n_X,n_{\rm tot})$. 
%Therefore, if $G(m_X,k_{\rm tot},n_{\rm tot})$ is a decreasing function of $k_{\rm tot}$, 
%we can use the same argument to $f_{\rm BI}(k_X)$. 
Since $G(n_X;k_{\rm tot},n_{\rm tot})$ is a decreasing function of $k_{\rm tot}$, 
by using an argument similar to the one leading to Eq.~(\ref{opa}), we have  
\begin{equation}
\sum_{n_X;~k_{\rm tot} > f_{\rm opt}^{(k_X=0)}(n_X,n_{\rm tot})} 
{\rm HG}(0;n_X,k_{\rm tot},n_{\rm tot}){\rm BI}(n_X;n_{\rm tot},p_X)
\leq \epsilon_{\rm PE}. 
\end{equation} 
This is identical to Eq.~(\ref{niko}) 
since $k_X < k_{\rm tot}-f(k_X,n_X,n_{\rm tot})$ is never satisfied for $k_X\geq 1$. 
The key length when $k_X=0$ was observed is then given by 
\begin{equation} 
l^{(\rm opt)}:= n_Z(1-h\left (\frac{f_{\rm opt}^{(k_X=0)}(n_X,n_{\rm tot})}{n_Z}\right))
-{\rm log_2} \frac{2}{\epsilon_{\rm PA}} - \lambda_{\rm EC}(\epsilon_{\rm c}).
\label{optkeyrate}
\end{equation}

\begin{figure}[t]
  \begin{center}
  \raisebox{0mm}
  {\includegraphics[width=80mm]{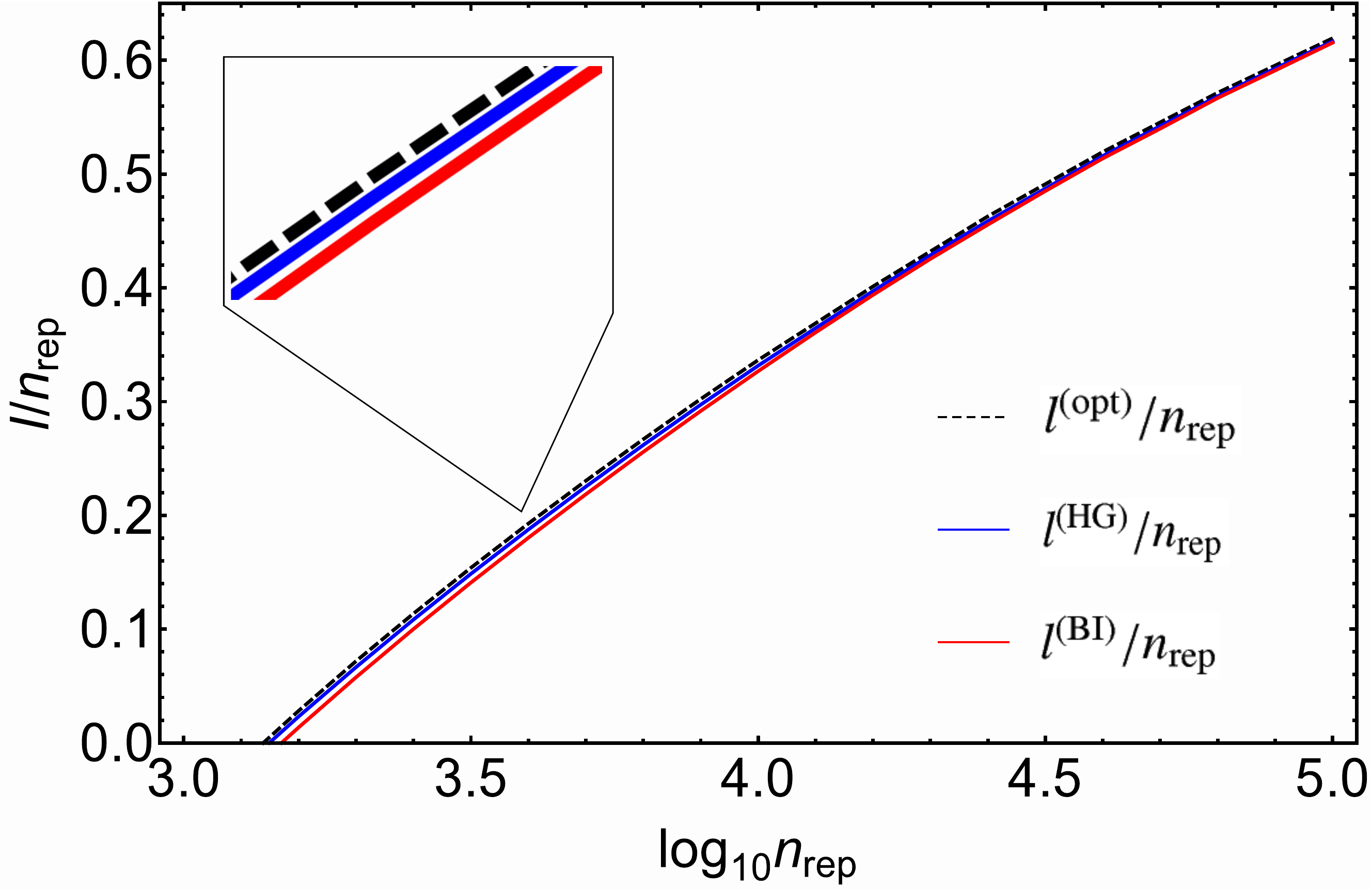}}
   \end{center}
   \caption{Secure key ratio of the qubit-based BB84 protocol 
   to the asymptotic limit 
    as a function of total rounds of 
   the protocol $n_{\rm rep}$. 
   We assume no errors ($k_X=0$) and no loss 
   ($n_{\rm tot}=n_{\rm rep}$). 
   The security parameters are set to $\epsilon_{\rm c}=10^{-15}$ and 
   $\epsilon_{\rm s}=10^{-10}$. 
The top, middle and bottom curves represent the ratios 
$l^{(\rm opt)}/n_{\rm rep}$, $l^{(\rm HG)}/n_{\rm rep}$ (method with simple random sampling) 
and $l^{(\rm BI)}/n_{\rm rep}$ (Bernoulli-sampling method), respectively. 
In the limit of $n_{\rm rep}\to \infty$,  each curve converges to $l/n_{\rm rep}=1$. 
   }
   \label{3comp}
\end{figure}

In Fig.~\ref{3comp}, 
we show the secure key ratios to the asymptotic case 
$l^{(\rm BI)}/n_{\rm rep}$, $l^{(\rm HG)}/n_{\rm rep}$ and $l^{(\rm opt)}/n_{\rm rep}$ 
as functions of total rounds of the protocol $n_{\rm rep}$. 
For each $n_{\rm rep}$, the value of $\tilde{p}_X$ was optimized to maximize the key length. 
In the limit of $n_{\rm rep}\to \infty$, each curve converges to $l/n_{\rm rep}=1$. 
The security parameters are set to 
$\epsilon_{\rm c}=10^{-15}$ and 
   $\epsilon_{\rm s}=10^{-10}$, $\epsilon_{\rm PE}=1/4\times 10^{-20}$ and 
   $\epsilon_{\rm PA}=1/4\times 10^{-20}$. We see that 
although the key rate $l^{(\rm opt)}$ is the best, 
the three methods achieve almost the same key length.

\section{Analysis for WCP-based protocol}\label{WCPBB84}
Here, we 
apply the analyses introduced in the previous section to the protocols 
using weak coherent pulses (WCP). 
We consider the WCP-based BB84 protocol in the subsections A and B, and 
move to the DQPS protocol in subsection C. 

\subsection{The WCP-BB84 protocol}\label{WCPBB84}
Similarly to the ideal qubit-based protocol, 
the WCP-BB84 protocol 
also follows the procedures described in Sec.~\ref{protocol}, 
but the latter assumes more general light sources and measurement apparatuses. 
We prove the security of  the WCP-BB84 protocol based on that of qubit-based BB84 protocol 
combined with GLLP's tagging idea \cite{2004GLLP}. 
We impose the following assumptions on Alice and Bob's devices. 
For Alice's light source, we assume that the four states 
$\hat{\rho}_{W,a}$ 
with $W\in \{Z,X\}$ and $a\in \{0,1\}$ 
are written as 
\begin{equation}
\hat{\rho}_{W,a}=(1-r_{\rm tag})\hat{\rho}_{W,a,\rm unt}\oplus r_{\rm tag}\hat{\rho}_{W,a,\rm tag}.
\label{joutai}
\end{equation}
For $\{\hat{\rho}_{W,a,\rm unt}\}$, we assume that 
there is a basis-independent state $\hat{\chi}_{\rm unt}$ on the system $AS$ satisfying
\begin{equation} 
{\rm tr}_A\left((\ket{a_W}\bra{a_W}_A\otimes \hat{\mathbbm{1}}_S)\hat{\chi}_{\rm unt}\right)
=\frac{1}{2}\hat{\rho}_{W,a,\rm unt}.
\label{Ma}
\end{equation}
%which indicates that the measurement to obtain phase error on $Z$-labeled incident is identical to 
%the that to obtain bit error on $X$-labeled incident. 
%For the security proof with entangement distillation, Eq.~(\ref{Ma}) is satisfied with 
%$\hat{\chi}=\ket{\Phi^+}\bra{\Phi^+}$. 
We place no restriction on the states $\{\hat{\rho}_{W,a,\rm tag}\}$. 
We may still find a state $\hat{\chi}_{W,\rm tag}$ on the system $AS$ such that 
\begin{equation}
{\rm tr}_A\left((\ket{a_W}\bra{a_W}_A\otimes \hat{\mathbbm{1}}_S)\hat{\chi}_{W,\rm tag}\right)
=\frac{1}{2}\hat{\rho}_{W,a,\rm tag},
\label{Xionfeng}
\end{equation}
but $\hat{\chi}_{W,\rm tag}$ depends on a selected basis $W\in\{Z,X\}$. 
The form of Eq.~(\ref{joutai}) allows an interpretation that each round is classified as 
either tagged or untagged \cite{2004GLLP}. The density operator $\hat{\rho}_{W,a,\rm unt}$ is the state 
of Alice's untagged signal which may generate 
a secure key and 
$\hat{\rho}_{W,a,\rm tag}$ is that of her tagged signal which is considered to be totally insecure. 
Equation (\ref{Ma}) indicates that Alice's basis choice can be postponed until Bob receives the system $S$ 
for untagged rounds. 
Eqs.~(\ref{joutai}) and (\ref{Ma}) are realized, for example,  
if Alice uses a laser emitting an ideally polarized coherent pulse with mean photon number 
$\mu$ and randomizes its optical phase. 
In this case, $\hat{\rho}_{W,a,\rm unt}$ and $\hat{\rho}_{W,a,\rm tag}$ are written as 
\begin{align}
(1-r_{\rm tag})
\hat{\rho}_{W,a,\rm unt}&=
e^{-\mu}\ket{0_{W,a}}\bra{0_{W,a}}
+\mu e^{-\mu}\ket{1_{W,a}}\bra{1_{W,a}}
\\
r_{\rm tag}
\hat{\rho}_{W,a,\rm tag}&=
e^{-\mu}\sum_{m=2}^{\infty}\frac{\mu^m}{m!}\ket{m_{W,a}}\bra{m_{W,a}}
\end{align}
where 
$\ket{m_{W,a}}$ is an $m$-photon state with a basis $W$ and a bit $a$, and 
\begin{equation}
r_{\rm tag}=1-e^{-\mu}-\mu e^{-\mu}
\label{taguttagu}
\end{equation}
 represents the probability that Alice emits 
two or more photons. 
Our proof does not depend on the specific model such as coherent light source, 
but depends only on Eqs.~(\ref{joutai}) and (\ref{Ma}). 
%(For example, $\ket{H}\bra{H}+\ket{V}\bra{V}=\ket{D}\bra{D}+\ket{\bar{D}}\bra{\bar{D}}$ for polarization case). 

For Bob's measurement apparatus, we impose either of the following two assumptions. 
\\(i) The probability of detecting a signal at Bob's receiver is independent of 
his basis choice. \\
(ii) 
The measurement of an input signal on the system $S$ 
is replaced by an ideal single-photon measurement on the system $B$ preceding by 
 a squashing operation \cite{2008Tsurumaru,2008Lutkenhaus}. \\
The condition (i), which is weaker than condition (ii), allows us to use the security proof with 
complementarity \cite{2009Koashi} and uncertainty principle \cite{2012Tomamichel}. 
The condition (ii) 
validates the use of the security proof with entanglement distillation \cite{2000Shor}. 
For the WCP-BB84 protocol, both conditions are satisfied 
if we assume the following model for Bob's apparatus: 
Bob actively chooses the basis, and uses two threshold detectors corresponding to 
the measurement result ``0'' and ``1'' after a polarization beam splitter. 
He assigns random bit if both detectors report their detections. 
In addition, the inefficiency and dark countings of the detectors are 
allowed as long as they are equivalently represented by an
absorber and a stray photon source placed in front of Bob's apparatus. 

For the proof with entanglement distillation, 
we use Eqs.~(\ref{Ma}), (\ref{Xionfeng}) and the assumption (ii) of Bob's apparatus to convert 
the actual protocol equivalently to a protocol in which Alice and Bob make ideal measurements on 
the qubit systems $A$ and $B$. 
Then a phase error is defined in the same way to the ideal BB84 protocol, namely, 
an error between Alice's and Bob's outcomes of ideal $X$-basis measurements ($\{\ket{0_X},\ket{1_X}\}$) 
on a $Z$-labeled round. 
As for the proof with complementarity, 
phase error in a $Z$-labeled round is defined as an error occurring when 
Alice makes an ideal $X$-basis measurement 
on the system $A$ and 
Bob makes the actual $X$-basis measurement on the system $S$ 
(the measurement conducted on $X$-labeled rounds in the actual protocol).

The secure key length formulated in 
Eqs.~(\ref{f})-(\ref{keyrate}) for the ideal protocol can be adapted to the WCP protocol through the tagging idea. 
Let $k_{\rm ph,unt}$ be the total number of phase errors on the untagged $Z$-labeled rounds. 
Let $n_{Z,\rm unt}$ be the number of untagged $Z$-labeled rounds. 
Since each round can be classified as tagged or untagged, 
$n_{Z,\rm unt}$ is a well-defined random variable in the actual protocol. 
Suppose that an upper bound of $k_{\rm ph,unt}$ is given as a function of 
$k_X$, $n_X$, $n_{\rm tot}$ and $n_{Z,\rm unt}$: 
\begin{equation}
{\rm Pr}(k_{\rm ph,unt} > f(k_X,n_X,n_{\rm tot},n_{Z,\rm unt}))\leq \epsilon_{\rm PE}.
\label{chaos}
\end{equation}
According to the tagging idea,  
the final key is $\epsilon_{\rm s}$-secret if 
\begin{align} 
n_{\rm fin}\leq 
n_{Z,\rm unt}(1-h\left (\frac{ f(k_X,n_X,n_{\rm tot},n_{Z,\rm unt})}
{n_{Z,\rm unt}}\right)) 
-{\rm log_2} \frac{2}{\epsilon_{\rm PA}} - \lambda_{\rm EC}(\epsilon_{\rm c}) 
\label{keytag}
\end{align} 
is satisfied. 
In the practical situation, the exact value of $n_{Z,\rm unt}$ is not available, 
and hence it is impossible to satisfy Eq.~(\ref{keytag}) with certainty. 
Instead, we allow a small error probability $\epsilon_{Z,\rm unt}$. 
Suppose that there is a probabilistic lower bound 
$\munderbar{n}_{Z,\rm unt}$ 
 which satisfies 
\begin{equation}
{\rm Pr}(n_{Z,\rm unt} < \munderbar{n}_{Z,\rm unt})\leq \epsilon_{Z,\rm unt}.
\label{kagen}
\end{equation}
A key length as a function of observed values $k_X,n_X,n_{\rm tot}$ 
is given by minimizing the right-hand side of Eq.~(\ref{keytag}) in the range of 
$n_{Z,\rm unt}\geq \munderbar{n}_{Z,\rm unt}$.

%Combined with Eqs.~(\ref{keytag}) and (\ref{kagen}), 
%by setting  
%\begin{equation}
%\epsilon_{\rm sec}=\epsilon_{\rm c}+\sqrt{2}\sqrt{\epsilon_{\rm PE}+\epsilon_{\rm PA}}+\epsilon_{Z,\rm unt}
%+\epsilon_{X,\rm unt},
%\label{rice}
%\end{equation}
%the protocol is $\epsilon_{\rm sec}$-secure 
%if 
%\begin{align} 
%l\leq \mathop{\rm min}_{n_{Z,\rm unt}\geq \munderbar{n}_{Z,\rm unt},n_{X,\rm unt}\geq \munderbar{n}_{X,\rm unt}}
%n_{Z,\rm unt}(1-h\left (\frac{f(k_{X,\rm unt},n_{X,\rm unt},n_{\rm tot,unt})}
%{n_{Z,\rm unt}}\right)) \nonumber \\
%~~~~~~~~~~~~~~~~~~~~~-\lambda_{\rm EC}-
%{\rm log_2} \frac{1}{\epsilon_{\rm c} \epsilon_{\rm PA}}. 
%\label{keycommon}
%\end{align} 
%The above key length formula holds for general analysis. 
%If we apply the approach with binomial distribution which is introduced in the section~\ref{bounds}, 
%we have 
%\begin{equation}
%{\rm Pr}(k_{\rm ph,unt}\geq f(k_X))\leq \epsilon_{\rm PE},
%\label{ajinomoto}
%\end{equation}
%which is shown in later. 

Under the assumptions for the source and measurement apparatus, 
the basic distributions used in the previous section, 
Eqs.~(\ref{BIk}) and (\ref{HG}), are still valid if we confine ourselves to the untagged rounds. 
Although the fact may be intuitively obvious for the 
WCP-BB84 protocol, 
here we give its mathematical justification since it helps when we treat a less intuitive protocol 
in Subsection \ref{appDQPS}. We define 
a set of integers labeling the rounds in the protocol as $\mathcal{N}_{\rm rep}:= \{1,2,....n_{\rm rep}\}$. 
As subsets of  $\mathcal{N}_{\rm rep}$, let us define 
the set of the integers labeling the rounds where Alice (Bob) chooses $X$ basis as $\mathcal{X}_A$ 
($\mathcal{X}_B$) regardless of detection. 
Define those 
labeling the untagged and detected rounds as $\mathcal{N}_{\rm unt}$. 
%We can not directly apply the analysis for the qubit-based protocol to the WCP protocol 
%because Eve can tell the difference of tagged state on $Z$ basis and $X$ basis and 
%errors are not necessarily distributed at random. 
%On the other hand, Eq.~(\ref{nozaki}) holds for untagged incidents, 
%which allows us to use the ``source replacement'' (introduced in the qubit-protocol section) 
%that assumes preparation of a maximally entangled state. 
Let $\mathcal{K}_{\rm unt}$ be a subset of $\mathcal{N}_{\rm unt}$ labeling the rounds 
which have errors when Alice and Bob conduct virtual $X$-basis measurements regardless of their basis choice. 
For any subset $\mathcal{M}$, let 
$\overbar{\mathcal{M}}:=\mathcal{N}_{\rm rep}\setminus \mathcal{M}$. 
With these notations, 
$k_{\text{ph,unt}}= |\overbar{\mathcal{X}_A}\cap
			 \overbar{\mathcal{X}_B}
			 \cap\mathcal{K}_{\mathrm{unt}}|$ 
and  
$n_{\text{Z,unt}}=	|\overbar{\mathcal{X}_A}\cap
			 \overbar{\mathcal{X}_B}
			 \cap\mathcal{N}_{\mathrm{unt}}|$. 
We define other random variables as follows:
$k_{X,\text{unt}}:=|\mathcal{X}_A\cap\mathcal{X}_B\cap\mathcal{K}_{\mathrm{unt}}|$, 
$n_{X,\text{unt}}:=|\mathcal{X}_A\cap\mathcal{X}_B\cap\mathcal{N}_{\mathrm{unt}}|$, 
$k_{\rm tot,unt}:=k_{X,\rm unt}+k_{\rm ph, unt}$ and 
$n_{\rm tot,unt}:=n_{X,\rm unt}+n_{Z,\rm unt}$. 
From the assumption of Alice's source Eq.~(\ref{Ma}) and the assumptions of 
Bob's receiver (i) or (ii), 
the choice of $\mathcal{X}_A\cap \mathcal{N}_{\rm unt}$ and 
$\mathcal{X}_B\cap \mathcal{N}_{\rm unt}$ can be postponed after $\mathcal{N}_{\rm unt}$ and 
 $\mathcal{K}_{\rm unt}$ are determined 
as far as untagged incidents are concerned.  
Then we have 
\begin{align}
 &\text{Pr}(\mathcal{X}_A\cap\mathcal{N}_{\mathrm{unt}}=\mathcal{M}_A,
 \mathcal{X}_B\cap\mathcal{N}_{\mathrm{unt}}=\mathcal{M}_B \mid \mathcal{K}_{\text{unt}},\mathcal{N}_{\text{unt}})
	\nonumber 
        \\ &= \Theta(\mathcal{M}_A,\mathcal{N}_{\rm unt})\Theta(\mathcal{M}_B,\mathcal{N}_{\rm unt})
 \label{gozilla}
\end{align}
for all $\mathcal{M}_A\subset \mathcal{N}_{\rm unt}$ and $\mathcal{M}_B\subset \mathcal{N}_{\rm unt}$, 
where we defined 
 \begin{equation}
 \Theta(\mathcal{M}_1,\mathcal{M}_2)=
 \tilde{p}_X^{|\mathcal{M}_1|} \tilde{p}_Z^{|\mathcal{M}_2\setminus \mathcal{M}_1|}. 
 \label{where we de}
 \end{equation} 
By simple calculation of the probability theory, we have 
\begin{equation}
{\rm Pr}(k_{X,\rm unt}\mid k_{\rm tot,unt},n_{\rm tot,unt})=
{\rm BI}(k_{X,\rm unt};k_{\rm tot,unt},p_X)\label{BIkunt}. 
\end{equation} 
and
\begin{align}
 {\rm Pr}(k_{X,\rm unt}\mid n_{X,\rm unt},k_{\rm tot,unt},n_{\rm tot,unt})=
 {\rm HG}(k_{X,\rm unt};n_{X,\rm unt},k_{\rm tot,unt},n_{\rm tot,unt}),\label{HGunt}
\end{align} 
which 
means that Eqs.~(\ref{BIk}) and (\ref{HG}) essentially hold true for the untagged rounds.

Now we derive a key rate formula for the WCP BB84 protocol based on Eq.~(\ref{BIkunt}), 
as was done with the Bernoulli-sampling method for the qubit-based protocol in Sec.~\ref{bounds}. 
First, we seek for $f(k_X,n_X,n_{\rm tot},n_{Z,\rm unt})$ which satisfies \eq{chaos}. 
Analogous to the derivation of Eq.~(\ref{ffbi}) from Eq.~(\ref{BIk}), 
 Eq.~(\ref{BIkunt}) leads to  
\begin{equation}
{{\rm Pr}(k_{\rm ph,unt} > f_{\rm BI}(k_{X,\rm unt})\mid k_{\rm tot,unt},n_{\rm tot,unt})\leq \epsilon_{\rm PE}} 
 \end{equation} 
for any $k_{\rm tot,unt}$ and $n_{\rm tot,unt}$, and hence we have  
\begin{equation}
{\rm Pr}(k_{\rm ph,unt} > f_{\rm BI}(k_{X,\rm unt}))\leq \epsilon_{\rm PE}.
\label{fwcpb}
\end{equation}
Since $k_{X,\rm unt}$ is not an observed value, 
 we use the obvious bound 
\begin{equation}
k_{X,\rm unt}\leq k_X \label{kx}. 
\end{equation}
Using the inequality 
\begin{align}
 &C_{\rm BI}(k_X+1;k_{\rm tot}+1,p_X)\nonumber \\
 =&~C_{\rm BI}(k_X;k_{\rm tot},p_X)+
(1-p_X){\rm BI}(k_X+1;k_{\rm tot},p_X)\nonumber \\
\geq&~C_{\rm BI}(k_X;k_{\rm tot},p_X)
\end{align}
 in Eq.~(\ref{fb}), 
we have $f_{\rm BI}(k_X)\leq f_{\rm BI}(k_X+1)$, implying that 
$f_{\rm BI}(k_X)$ is an increasing function. 
Hence, Eqs.~(\ref{fwcpb}) and (\ref{kx}) lead to
\begin{equation}
{\rm Pr}(k_{\rm ph,unt} > f_{\rm BI}(k_X))\leq \epsilon_{\rm PE},
\label{pata}
\end{equation}
which means that $f_{\rm BI}(k_X)$ fulfills \eq{chaos}.

Next, we determine $\munderbar{n}_{Z,\rm unt}$ which satisfies Eq.~(\ref{kagen}). 
%Since the upper bound of $k_{\rm ph,unt}$ is independent of $n_{Z,\rm unt}$ in Eq.~(\ref{pata}), 
%Eq.~(\ref{keytag}) implies that a lower bound of the secure key rate is obtained if 
%a lower bound of $n_{Z,\rm unt}$ is known. 
To determine a lower bound of $n_{Z,\rm unt}$, we consider an upper bound of
 $n_{Z,\rm tag}:=n_Z-n_{Z,\rm unt}$. 
Let $N_{Z,\rm tag}$ be the number of rounds where Alice chooses $Z$ basis, 
Bob chooses $Z$ basis and 
the light source emits a tagged signal (two photons or more). 
As those conditions are independent of each other as seen from Eq.~(\ref{joutai}), 
we have 
\begin{equation}
{\rm Pr}(N_{Z,\rm tag})={\rm BI}(N_{Z,\rm tag},n_{\rm rep},r_{\rm tag} \tilde{p}_Z^2). \label{ariz}
\end{equation}
Since $n_{Z,\rm tag}$ is the number of detected rounds among the $N_{Z,\rm tag}$ rounds, 
\begin{equation}
n_{Z,\rm tag}\leq N_{Z,\rm tag} \label{ona}
\end{equation}
holds. 
Eqs.~(\ref{ariz}) and (\ref{ona}) lead to 
\begin{equation}
{\rm Pr}(n_{Z,\rm tag} > n)\leq 1-C_{\rm BI}(n;n_{\rm rep}, r_{\rm tag}\tilde{p}_Z^2)
\label{hitori}
\end{equation}
for any $n$. 
Thus, we have 
\begin{equation}
{\rm Pr}(n_{Z,\rm tag}> g(r_{\rm tag}\tilde{p}_Z^2,\epsilon_{Z,\rm unt}))\leq \epsilon_{Z,\rm unt}
\label{gomasuri}
\end{equation}
where 
\begin{align}
g(x,y):={\rm min}
\left\{
n\Big|1-C_{\rm BI}(n;n_{\rm rep},x) 
\leq y \right \}. 
\label{g}
\end{align}
Let $\munderbar{n}_{Z,\rm unt}$ be 
\begin{equation}
\munderbar{n}_{Z,\rm unt}:=n_Z-g(r_{\rm tag}\tilde{p}_Z^2,\epsilon_{Z,\rm unt}).
\label{gadget}
\end{equation} 
By using $n_{Z,\rm tag}=n_Z-n_{Z,\rm unt}$, Eq.~(\ref{gomasuri}) leads to 
\begin{align}
{\rm Pr}(n_{Z,\rm unt} < \munderbar{n}_{Z,\rm unt}) \leq \epsilon_{Z,\rm unt}. \label{ubarnz}
\end{align}

Since $n_{Z,\rm unt}$ is known in principle in the actual protocol, the final state 
$\rho_{AE}^{\rm fin}$ is written as a direct sum of the part for 
$n_{Z,\rm unt} < \munderbar{n}_{Z,\rm unt}$ 
and the one for $n_{Z,\rm unt} \geq \munderbar{n}_{Z,\rm unt}$. 
Hence, combined with 
Eqs.~(\ref{keytag}), (\ref{pata}) and (\ref{ubarnz}), 
by setting  
\begin{equation}
\epsilon_{\rm s}=\sqrt{2}\sqrt{\epsilon_{\rm PE}+\epsilon_{\rm PA}}+\epsilon_{Z,\rm unt},
\label{gapao}
\end{equation}
the protocol is $\epsilon_{\rm c}$-correct and $\epsilon_{\rm s}$-secret 
if 
\begin{align} 
n_{\rm fin} \leq l_{\rm WCP}^{(\rm BI)}:=
\munderbar{n}_{Z,\rm unt}(1-h\left (\frac{f_{\rm BI}(k_X)}
{\munderbar{n}_{Z,\rm unt}}\right)) 
-{\rm log_2} \frac{2}{\epsilon_{\rm PA}} - \lambda_{\rm EC}(\epsilon_{\rm c}). 
\label{keywcpb}
\end{align} 
Together with Eqs.~(\ref{fb}), (\ref{cb}), (\ref{g}) and (\ref{gadget}), 
Eq.~(\ref{keywcpb}) constitutes the main result of Sec \ref{WCPBB84}.

For the purpose of comparison, here we also discuss what the key rate formula looks like 
if we start from Eq.~(\ref{HGunt}), based on simple random sampling. 
As we have derived Eq.~(\ref{ffhg}) from Eq.~(\ref{HG}), 
Eq.~(\ref{HGunt}) leads to 
\begin{equation}
{{\rm Pr}(k_{\rm ph,unt} > f_{\rm HG}(k_{X,\rm unt},n_{X,\rm unt},n_{\rm tot,unt})\mid k_{\rm tot,unt},n_{\rm tot,unt})\leq \epsilon_{\rm PE}}, 
 \end{equation} 
which, in turn, leads to  
\begin{equation}
{\rm Pr}(k_{\rm ph,unt} > f_{\rm HG}(k_{X,\rm unt},n_{X,\rm unt},n_{\rm tot,unt}))\leq \epsilon_{\rm PE}. 
\label{fwcphg}
\end{equation} 
Similarly to $f_{\rm BI}(k_X)$, we can prove that 
$f_{\rm HG}(k_X,n_X,n_{\rm tot})$ is an increasing function of $k_X$. 
Since $k_{X,\rm unt}$ is upper-bounded by Eq.~(\ref{kx}), 
 Eq.~(\ref{fwcphg}) leads to 
\begin{equation}
{\rm Pr}(k_{\rm ph,unt}> f_{\rm HG}(k_X,n_{X,\rm unt},n_{\rm tot,unt}))\leq \epsilon_{\rm PE}. 
\label{fwcphgg}
\end{equation}
In contrast to Eq.~(\ref{pata}), 
it requires an additional estimation process for $n_{X,\rm unt}$ to obtain 
$f_{\rm HG}(k_X,n_{X,\rm unt},n_{\rm tot,unt})$. A lower bound defined by 
$\munderbar{n}_{X,\rm unt}:=n_X-g(r_{\rm tag}\tilde{p}_X^2,\epsilon_{X,\rm unt})$ 
satisfies 
\begin{align}
 {\rm Pr}(n_{X,\rm unt} < \munderbar{n}_{X,\rm unt})\leq \epsilon_{X,\rm unt} \label{ubarnx}. 
\end{align}  
Combined with Eqs.~(\ref{keytag}), (\ref{fwcphgg}) and (\ref{ubarnx}),
by setting 
\begin{equation}
\epsilon_{\rm s}=\sqrt{2}\sqrt{\epsilon_{\rm PE}+\epsilon_{\rm PA}}+
\epsilon_{Z,\rm unt}+\epsilon_{X,\rm unt}, 
\end{equation} 
the protocol is $\epsilon_{\rm c}$-correct and $\epsilon_{\rm s}$-secret 
if 
\begin{align} 
n_{\rm fin} \leq~l_{\rm WCP}^{(\rm HG)}&:= \min_{n_{Z,\rm unt}\geq \munderbar{n}_{Z,\rm unt}} 
\xi(k_X,\munderbar{n}_{X,\rm unt},n_{Z,\rm unt}) 
\nonumber \\
\xi(k_X,\munderbar{n}_{X,\rm unt},n_{Z,\rm unt}) &:=
\tilde{\xi}(k_X,\munderbar{n}_{X,\rm unt},n_{Z,\rm unt})
-{\rm log_2} \frac{2}{\epsilon_{\rm PA}} - \lambda_{\rm EC}(\epsilon_{\rm c})
\nonumber  \\ 
\tilde{\xi}(k_X,\munderbar{n}_{X,\rm unt},n_{Z,\rm unt})& 
\nonumber \\ 
:=
n_{Z,\rm unt}&(1-h\left (\frac{f_{\rm HG}(k_X,\munderbar{n}_{X,\rm unt},\munderbar{n}_{X,\rm unt}+n_{Z,\rm unt})
}
{n_{Z,\rm unt}}\right)).
\label{koreya}
\end{align} 
The reason that the minimization of $n_{Z,\rm unt}$ appears is because 
$\tilde{\xi}(k_X,\munderbar{n}_{X,\rm unt},n_{Z,\rm unt})$ is not monotone-increasing function of $n_{Z,\rm unt}$. 
For example, with 
$\epsilon_{\rm PE}=1/16\times 10^{-20}$, we numerically confirmed 
that $\tilde{\xi}(0,25000,25318)\sim 24631$ 
%24630.8 
and $\tilde{\xi}(0,25000,25319)\sim 24623$.  
%24623.2
%Since $f_{\rm HG}(k_X,\munderbar{n}_{X,\rm unt},\munderbar{n}_{X,\rm unt}+n_{Z,\rm unt})$ is 
%an integer which is given by Eq.~(\ref{fhg}), 
% at some $n_{Z,\rm unt}$ we have 
% \begin{align}
% &f_{\rm HG}(k_X,\munderbar{n}_{X,\rm unt},\munderbar{n}_{X,\rm unt}+
% n_{Z,\rm unt}+1)
% \nonumber \\
% =&
% f_{\rm HG}(k_X,\munderbar{n}_{X,\rm unt},\munderbar{n}_{X,\rm unt}+
% n_{Z,\rm unt})+1.
% \end{align} 
% Since $f_{\rm HG}/n_{Z,\rm unt}< (f_{\rm HG}+1)/(n_{Z,\rm unt}+1)$ holds 
% for $f_{\rm HG}<n_{Z,\rm unt}$, 
%we may have $\xi(n_{Z,\rm unt})>\xi(n_{Z,\rm unt}+1)$,
This means that the protocol with final key length 
 $l=\xi(k_X,\munderbar{n}_{X,\rm unt},\munderbar{n}_{Z,\rm unt})$ is not necessarily secure. 
%In fact, we numberically confirmed an example of $\xi(n_{Z,\rm unt})>\xi(n_{Z,\rm unt}+1)$, 
%which is shown in the following numerical section.

As can be seen from the comparison between Eqs.~(\ref{keywcpb}) and (\ref{koreya}), 
the method with simple random sampling is much more complicated than the 
Bernoulli-sampling method, 
involving an additional estimated parameter and a minimization. 
Moreover, as shown in Sec.~\ref{exam}, it tends to give a key rate lower than the Bernoulli-sampling method, 
probably because of the use of pessimistic bound on $n_{X,\rm unt}$.

\begin{figure}[t]
  \begin{center}
  \vspace{1.5mm}
  \raisebox{-1mm}
  {\includegraphics[width=77mm]{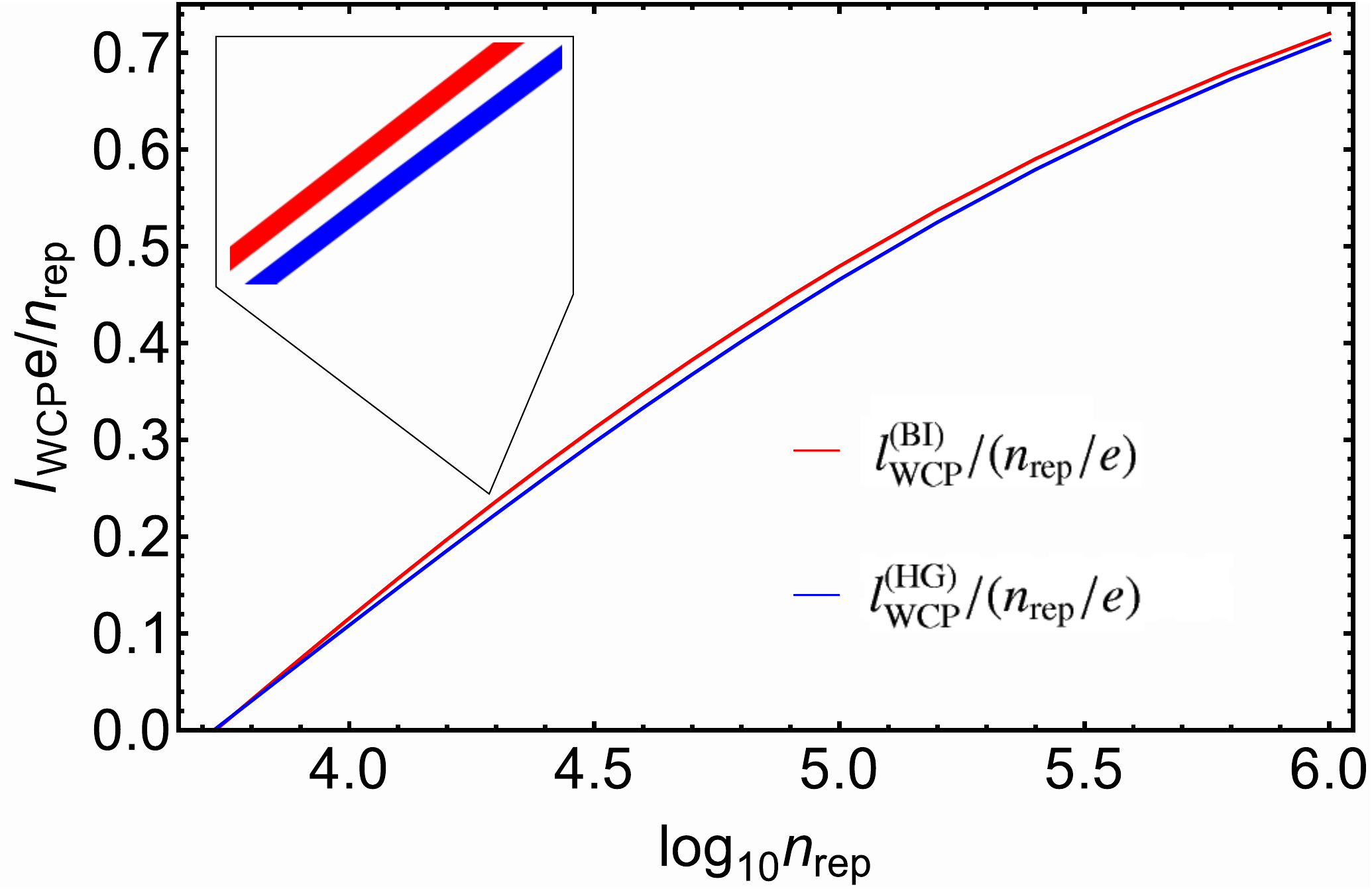}}
   \end{center}
   \caption{Comparison of estimation methods for the WCP-BB84 protocol. 
   Upper curve (Bernoulli-sampling method): Secure key ratio 
   to the asymptotic limit $l_{\rm WCP}^{(\rm BI)}/(n_{\rm rep}/e)$ 
    as a function of total rounds of 
   the protocol $n_{\rm rep}$. 
   Lower curve (method with simple random sampling): 
   An upper bound on the derived secure key ratio 
   $l_{\rm WCP}^{(\rm HG)}/(n_{\rm rep}/e)$. 
   We assume no error ($k_X=0$) and no loss 
   ($n_{\rm tot}=n_{\rm rep}(1-e^{-\mu})$). 
   The security parameters are set to $\epsilon_{\rm c}=10^{-15}$ and 
$\epsilon_{\rm s}=10^{-10}$. 
In the limit of $n_{\rm rep}\to \infty$,  each curve converges to $l_{\rm WCP}/(n_{\rm rep}/e)=1$. 
   }
   \label{wcpkeyrate}
\end{figure}

\begin{figure}[htbp]
  \begin{center}
  \raisebox{0mm}
  {\includegraphics[width=75mm]{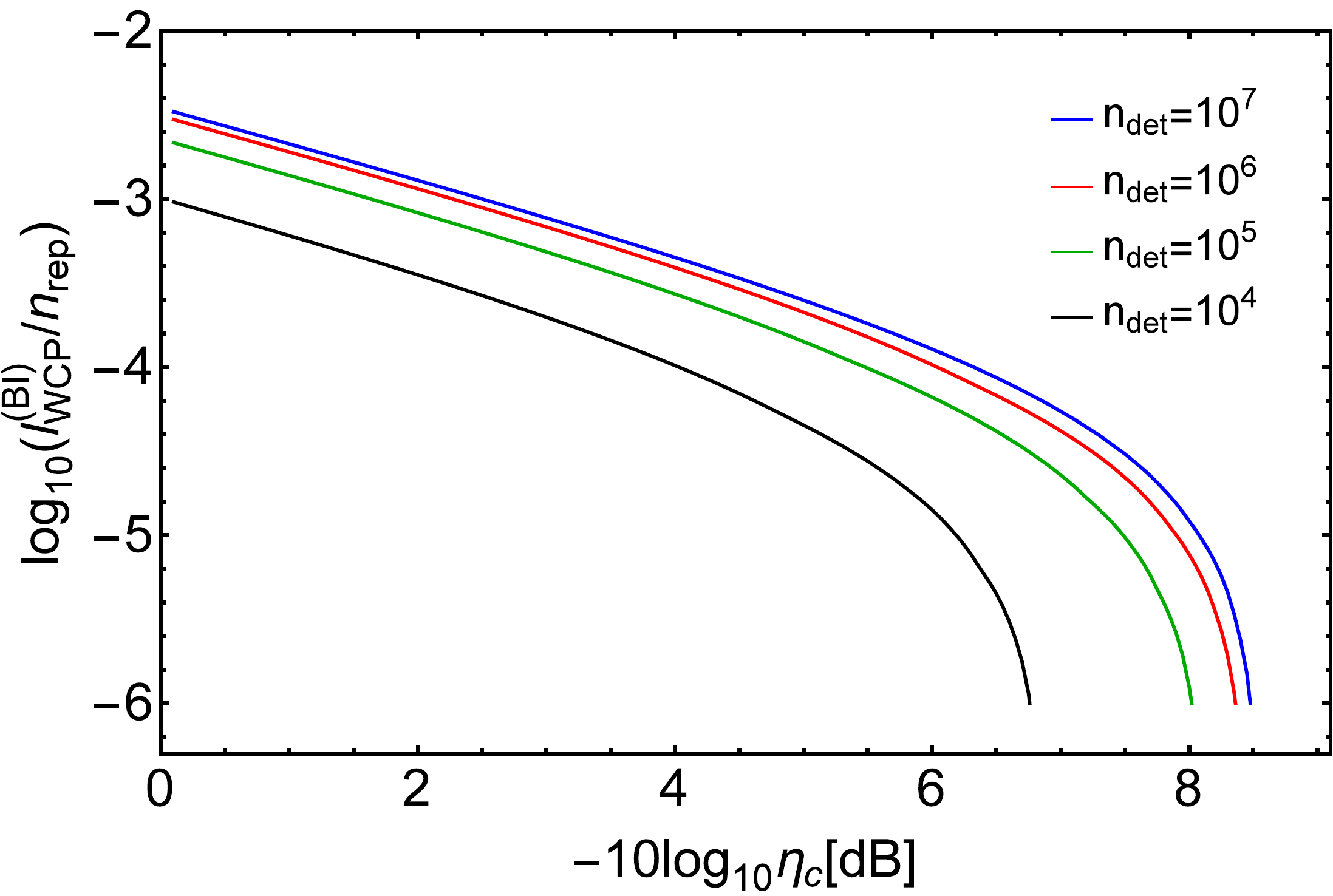}}
   \end{center}
   \caption{Secure key rate per signal of the WCP-BB84 protocol 
   $l_{\rm WCP}^{(\rm BI)}/n_{\rm rep}$ as a function of 
   channel transmission $\eta_c$. 
   The parameters are set to be the same as Ref.~\cite{2009Scarani}. 
Quantum efficiency of detectors: $\eta_d=0.1$.  
Dark count probability per pulse per detector: $p_{\rm dark}=10^{-5}$. 
Loss-independent bit error: $0.5\%$. 
Error correction cost: $\lambda_{\rm EC}(\epsilon_{\rm c})=1.05 h(E/Q)+{\rm log}_2 (1/\epsilon_{\rm c})$. 
The security parameters: $\epsilon_{\rm c}=10^{-10}$ and 
$\epsilon_{\rm s}=10^{-5}$. 
From the top to the bottom curve, the number of detected signals are 
$n_{\rm det}=10^7,10^6,10^5$ and $10^4$, respectively. 
The required number of detected signals to generate a final key is less than $10^4$, while it was 
$\sim 10^7$ in the previous result \cite{2009Scarani}. 
   }
   \label{wcpcompare}
\end{figure}

\subsection{Numerical examples}\label{exam}
Here, we show two examples of numerical calculation for the WCP-BB84 protocol. 
We assume that the light source emits a pulse whose 
 photon-number distribution is Poissonian with mean $\mu$, namely, \eq{taguttagu}. 
Like Fig.~\ref{3comp} for the ideal protocol, we first calculated the simplest case where 
no error is observed ($k_X=0$) and no loss occurs 
($n_{\rm tot}=n_{\rm rep}(1-e^{-\mu})$), which is shown in Fig.~\ref{wcpkeyrate}. 
The cost of error correction was set to 
$\lambda_{\rm EC}(\epsilon_{\rm c})={\rm log}_2 (1/\epsilon_{\rm c})$. 
We assumed $n_Z=n_{\rm tot}\tilde{p}_Z^2$ and $n_X=n_{\rm tot}\tilde{p}_X^2$. 
The values of $\tilde{p}_X$ and $\mu$ were optimized for each value of $n_{\rm rep}$. 
For calculation of $l_{\rm WCP}^{\rm (BI)}$, 
the security parameters were set to $\epsilon_{\rm c}=10^{-15}$, 
$\epsilon_{\rm s}=10^{-10}$, $\epsilon_{\rm PE}=1/16\times 10^{-20}$,  
$\epsilon_{\rm PA}=1/16\times 10^{-20}$ and $\epsilon_{Z,\rm unt}=1/2\times 10^{-10}$. 
The result is shown as the red curve in Fig.~\ref{wcpkeyrate}, 
where the key length Eq.~(\ref{keywcpb}) is normalized by the optimized asymptotic key rate
of $1/e$ per signal at $\mu=1$ and $\tilde{p}_X \to 0$.
We see that a final key can be extracted when the total rounds $n_{\rm rep}$ is 
more than $\sim 10^{3.7}$ while the threshold is 
$n_{\rm rep}\sim 10^{3.2}$ for the ideal protocol using the same parameters (see also Fig.~\ref{3comp}). 
For comparison, we also calculated the value of 
$\xi(k_X,\munderbar{n}_{X,\rm unt},\munderbar{n}_{Z,\rm unt})/(n_{\rm rep}/e)$ under the same condition, 
which is shown as the blue curve in Fig.~\ref{wcpkeyrate}. 
The security parameters were the same as the red curve, except for 
$\epsilon_{Z,\rm unt}=\epsilon_{X,\rm unt}=1/4\times 10^{-10}$. 
The quantity $\xi(k_X,\munderbar{n}_{X,\rm unt},\munderbar{n}_{Z,\rm unt})$ is an upper bound of 
$l_{\rm WCP}^{(\rm HG)}$ derived in Eq.~(\ref{koreya}). 
The figure shows that the key length $l_{\rm WCP}^{(\rm BI)}$ 
from Bernoulli sampling is higher than $l_{\rm WCP}^{(\rm HG)}$ from simple random sampling. 
A possible reason is that the estimation of 
$\munderbar{n}_{X,\rm unt}$, 
which is a pessimistic bound of $n_{X,\rm unt}$, 
is not required in determining $f_{\rm BI}(k_X)$. 

In Fig.~\ref{wcpcompare}, 
we show a result in more practical situations based on Eq.~(\ref{keywcpb}) to 
make comparison to the previous 
finite-key analysis for the WCP-BB84 protocol \cite{2009Scarani}. 
%As far as we know, this work is the only one which treats with finite-key analysis for the WCP-BB84 protocol. 
The figure shows 
the dependence of secure key rate $l_{\rm WCP}^{(\rm BI)}/n_{\rm rep}$ on 
the channel transmission $\eta_c$. 
In each curve, 
the number of Bob's detected signals $n_{\rm det}$ is fixed as $n_{\rm det}=10^4,10^5,10^6 $ and $10^7$. 
The parameters were chosen to be the same as \cite{2009Scarani}: 
Quantum efficiency of both detectors is $\eta_d=0.1$ and  
a dark count probability per pulse is $p_{\rm dark}=10^{-5}$ per detector. 
In addition to errors from dark counts, there is a $0.5\%$ loss-independent bit error. 
The security parameters were set to $\epsilon_{\rm c}=10^{-10}$, $\epsilon_{\rm s}=10^{-5}$, 
$\epsilon_{\rm PE}=1/16\times 10^{-10}$, $\epsilon_{\rm PA}=1/16\times 10^{-10}$, and 
$\epsilon_{Z,\rm unt}=1/2 \times 10^{-5}$. 
Total transmission rate is $Q=1-(1-2p_{\rm dark})e^{-\mu \eta_c \eta_d}$, and  
error rate is given by $E/Q$ where 
$E=0.005(1-e^{-\mu \eta_c \eta_d}) +p_{\rm dark}e^{-\mu \eta_c \eta_d}$. 
Based on the parameters above, 
we assume $\lambda_{\rm EC}(\epsilon_{\rm c})=1.05 h(E/Q)+{\rm log}_2 (1/\epsilon_{\rm c})$, 
$n_{\rm rep}=n_{\rm det}/Q$, $n_Z=n_{\rm det} \tilde{p}_Z^2$, 
 $n_X=n_{\rm det} \tilde{p}_X^2$ and $k_X=n_X E/Q$. 
To save the computation time, 
we used Chernoff bound \cite{1952Chernoff} 
\begin{align}
C_{\rm BI}(k_X;k_{\rm tot},p_X)\leq 
D\left(\frac{k_X}{k_{\rm tot}}, k_{\rm tot},p_X\right)
\label{doitu}
\end{align}
for $(k_X,k_{\rm tot},p_X)$ satisfying $k_X\leq k_{\rm tot}p_X$, 
where 
\begin{equation}
D(x,y,z):=\left(\Bigg(\frac{z}{x}\Bigg)^x \Bigg(\frac{1-z}{1-x}\Bigg)^{1-x} \right)^y.
\end{equation}
In Fig.~\ref{wcpcompare}, 
we see that a key can be extracted even when $n_{\rm det}= 10^4$.  
This is a significant improvement from the result of \cite{2009Scarani}, in which the required number of detected signals 
to generate a final key is  $n_{\rm det}\sim 10^7$.

\subsection{The DQPS protocol}\label{appDQPS}
In this section, we conduct finite-key analysis of the DQPS protocol based on the property 
of binomial distribution Eq.~(\ref{BIkunt}). 
The security of the DQPS protocol was recently proved in the asymptotic limit \cite{2016Kawakami}. 
The DQPS protocol 
uses encoding on four relative phases $\{0,\frac{\pi}{2},\pi,\frac{3\pi}{2}\}$ 
between neighboring pulses in a pulse train of fixed length $L$. 
The DQPS protocol has essentially the same setup as the BB84 protocol with phase encoding 
(PE-BB84 protocol), which can be regarded as the DQPS protocol with $L=2$. 
In Ref.~\cite{2016Kawakami}, we showed that the secure key rate of the DQPS protocol is 
8/3 as high as that of the PE-BB84 protocol in the asymptotic limit. 
However, since the security proof is not so straightforward as that of the BB84 protocol, 
it is not trivial whether the advantage of the DQPS protocol over the PE-BB84 protocol still holds 
considering the statistical fluctuations in the finite-key case. 
This motivates us to conduct finite-key analysis for the DQPS protocol 
by using the Bernoulli-sampling method proposed in this work. 

\begin{figure}[t]
  \begin{center}
  \raisebox{0mm}
  {\includegraphics[width=87mm]{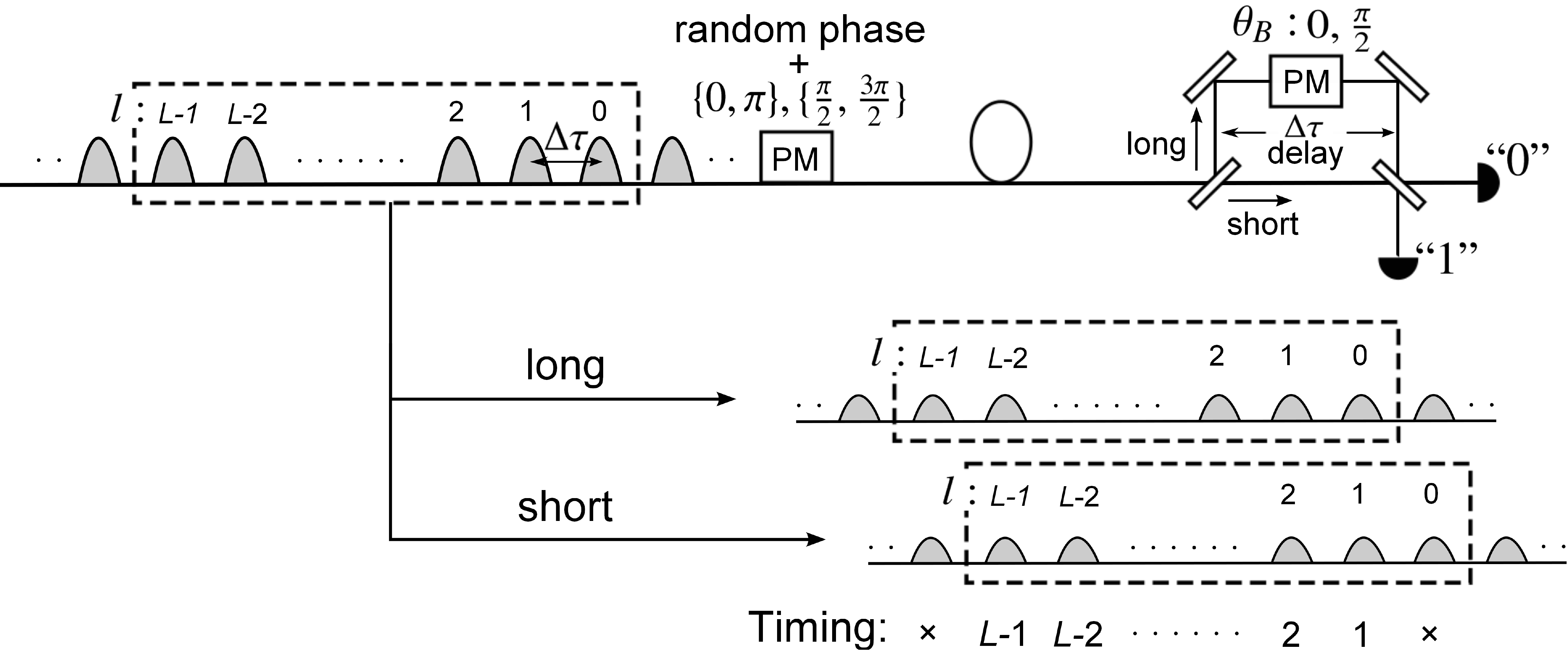}}
  \end{center}
  \begin{center}
  \caption{Setup for the $L$-pulse DQPS protocol. 
  %At Alice's site, pulse trains are generated by a laser 
%  followed by phase randomization as well as phase modulation (PM) 
%  
  The protocol regards a train of $L$ pulses as a block, 
and the working basis is chosen for each block. 
At Alice's site, pulses are modulated 
with phase $\{0,\pi,\frac{\pi}{2},\frac{3\pi}{2}\}$ according to her random bit and basis choice. 
The randomization of the overall optical phase is also done for each block of $L$ pulses. 
  At Bob's site, 
  each pulse train is fed to a delayed Mach-Zehnder interferometer with phase shift 
   0 or $\frac{\pi}{2}$ according to his basis choice.  
   %The trains leaving the interferometer are measured by two photon detectors corresponding to bit 
%   values ``0'' and ``1''.  
   Valid timings of detection are labeled by integers $1,2,..,L-1$, 
   according to the index of the pulse from the short arm of 
   the interferometer. Detection from  
   interference between pulses from different blocks is regarded as invalid and ignored. 
    \label{setupfig}
    }
   \end{center}
\end{figure}

The overview of the DQPS protocol is shown in Fig.~\ref{setupfig}. 
The precise description of the protocol and physical assumptions for the security proof is given in Appendix.  
In the protocol, 
$Z$ basis is chosen with probability $\tilde{p}_Z$ to generate keys and $X$ basis 
is chosen with probability $\tilde{p}_X$ for leak monitoring. 
Relative phases between adjacent pulses are modulated by $\{0,\pi\}$ for $Z$ basis 
and $\{\frac{\pi}{2},\frac{3\pi}{2}\}$ for $X$ basis. 
The protocol regards $L$-successive pulses as a block, and at most one key bit is extracted from each block. 
The randomization of the optical phase is conducted to the whole block, and a basis is also 
chosen for each block. 
Bob's receiver is composed of delayed interferometer with its delay being equal to the interval 
$\Delta \tau$ of adjacent pulses. 
The longer arm of the interferometer incurs phase shift 0 ($Z$-basis) or $\pi/2$ ($X$-basis),
which are chosen with probability $\tilde{p}_Z$ and $\tilde{p}_X$, respectively. 
After the interferometer, the pulses are measured by two photon detectors corresponding to bit values ``0'' and ``1''. 
If there is a detection from the superposition of the $l$-th and the $(l-1)$-th original pulses, 
we call it valid detection at $l$-th timing ($1\leq l\leq L-1$). 
An interference between different blocks at Bob's receiver is invalid and does not contribute to a key, 
which means that $1/L$ of the whole detection events must be discarded. 
This is the origin of 
the advantage of the DQPS protocol over the PE-BB84 protocol, which is regarded 
as the DQPS protocol with $L=2$.

In the DQPS protocol, 
application of the tagging idea is not straightforward since the 
 chain of coherence among successive pulses 
 prohibits us from defining the total photon number in neighboring two pulses. 
As a result, the conventional definition of tagging based on the emitted photon number cannot be applied here. 
In \cite{2016Kawakami}, 
we proposed an alternative approach to define the photon number indirectly through 
Bob's detection timing $j$ and Alice's 
measurement result on her qubits, which enables us to 
assume that each round in the protocol is classified as either tagged or untagged. 
As a result, the variables $k_{\rm ph,unt}$ and $n_{Z,\rm unt}$ can be defined in the same way as 
in the WCP-BB84 protocol, and the argument up to \eq{kagen} holds for the DQPS protocol as well. 
The remaining tasks are to find a function $f$ satisfying \eq{chaos} and to find a bound 
$\munderbar{n}_{Z,\rm unt}$ satisfying \eq{kagen}, 
both of which require slightly different approaches from the WCP-BB84 protocol.

Since our tagging definition for the DQPS protocol involves Bob's detection timing $j$, 
we cannot decompose the emitted states as in \eq{joutai}. 
Hence we need to justify \eq{gozilla} without using \eq{Ma}. 
This was essentially done in Ref.~\cite{2016Kawakami}, 
which proved, 
in the notation of the present paper~
\footnote
{The argument in Ref.~\cite{2016Kawakami} represented various events in $n_{\rm rep}$ rounds through 
strings $\bm{c},\bm{a},\bm{b},\bm{j}$ and $\bm{t}$ of length $n_{\rm rep}$, and 
proved that their joint probability distribution takes a form 
${\rm Pr}(\bm{c},\bm{a},\bm{b},\bm{j},\bm{t})
={\rm Pr}(\bm{c})\beta(g_{\bm{t},\bm{j}}(\bm{c}),\bm{a},\bm{b},\bm{j},\bm{t})$ 
as stated after Eq.(A3) of Ref.~\cite{2016Kawakami}. 
In rewriting it to Eq.~(\ref{didie}) of the main text, we used the following facts. 
There is a one-to-one correspondence between $\mathcal{X}_A$ and $\bm{c}$. 
$\mathcal{N}_{\rm unt}$ and $\mathcal{K}_{\rm unt}$ are functions of $\bm{a},\bm{b},\bm{j}$ and $\bm{t}$. 
$g_{\bm{t},\bm{j}}(\bm{c})$ is a function of 
$\{\mathcal{X}_A\cap \overbar{\mathcal{N}_{\rm unt}},\mathcal{N}_{\rm unt}\}$. 
%$\mathcal{X}_A\cap(\mathcal{N}_{\mathrm{rep}}\setminus\mathcal{N}_{\mathrm{unt}})$ is uniquely determined by 
%$g_{\bm{t},\bm{j}}(\bm{c})$, 
%$\mathcal{N}_{\text{unt}}$ and $\mathcal{K}_{\text{unt}}$ are 
%uniquely determined by $\bm{a},\bm{b},\bm{j},\bm{t}$.  
%The relation between $\beta'$ and $\beta$ is 
%\begin{equation}
% \beta'(\mathcal{X}_A\cap\overbar{\mathcal{N}_{\rm unt}},\mathcal{K}_{\text{unt}},\mathcal{N}_{\text{unt}})
%	=\sum_{\bm{a},\bm{b},\bm{j},\bm{t}}\beta(g_{\bm{t},\bm{j}}(\bm{c}),\bm{a},\bm{b},\bm{j},\bm{t})
%\end{equation}
%where the sum over $\bm{a},\bm{b},\bm{j},\bm{t}$ is under the fixed 
% $\mathcal{N}_{\text{unt}}$ and $\mathcal{K}_{\text{unt}}$. 
%The probability distribution of $\mathcal{X}_B$ was treated independently which is  
%represented by $p(\mathcal{X}_B)$ in this work. 
}, 
that the joint probability of 
$\mathcal{X}_A$, $\mathcal{K}_{\rm unt}$ and $\mathcal{N}_{\rm unt}$ is written in the following form: 
%%%%%%%%%%%%%%%%%
\begin{align}
 &\Pr(\mathcal{X}_A,\mathcal{K}_{\text{unt}},\mathcal{N}_{\text{unt}})
 \nonumber 
 \\
&=
	\Theta(\mathcal{X}_A,\mathcal{N}_{\rm rep})
        \beta'(\mathcal{X}_A\cap \overbar{\mathcal{N}_{\rm unt}},\mathcal{K}_{\text{unt}},\mathcal{N}_{\text{unt}}).
\label{didie}
\end{align}
Since 
$\Theta(\mathcal{M},\mathcal{N}_{\rm rep})$ 
defined in \eq{where we de} satisfies 
\begin{equation}
\Theta(\mathcal{M},\mathcal{N}_{\rm rep})=
\Theta(\mathcal{M}\cap \mathcal{N}_{\rm unt},\mathcal{N}_{\rm unt})
\Theta(\mathcal{M}\cap \overbar{\mathcal{N}_{\rm unt}},\overbar{\mathcal{N}_{\rm unt}})
\end{equation}
for any $\mathcal{M}\subset \mathcal{N}_{\rm rep}$, from Eq.~(\ref{didie}) we have 
\begin{align}
 &\Pr(\mathcal{X}_A\cap\mathcal{N}_{\mathrm{unt}}=\mathcal{M}_A
  \mid \mathcal{K}_{\text{unt}},\mathcal{N}_{\text{unt}}) 
 \nonumber \\ &=
		\Theta(\mathcal{M}_A,\mathcal{N}_{\rm unt})
		\gamma(\mathcal{K}_{\text{unt}},\mathcal{N}_{\text{unt}})
\label{ibara}
\end{align}
for any $\mathcal{M}_A \subset \mathcal{N}_{\rm unt}$, where
\begin{align}
 &\gamma(\mathcal{K}_{\text{unt}},\mathcal{N}_{\text{unt}}) 
 \nonumber \\ &:=
	\frac{
        \sum_{\mathcal{M}'_A\subset \overbar{\mathcal{N}_{\rm unt}}} 
        \Theta(\mathcal{M}'_A,\overbar{\mathcal{N}_{\rm unt}})
        \beta'(\mathcal{M}'_A,\mathcal{K}_{\text{unt}},\mathcal{N}_{\text{unt}})}
        {\Pr(\mathcal{K}_{\text{unt}},\mathcal{N}_{\text{unt}})}.
\end{align}
Since the sum of 
$\Theta(\mathcal{M}_A,\mathcal{N}_{\rm unt})$ over $\mathcal{M}_A$ is unity, 
Eq.~(\ref{ibara}) leads to 
 $\gamma(\mathcal{K}_{\text{unt}},\mathcal{N}_{\text{unt}})=1$. 
Thus, we have 
\begin{align}
 &\text{Pr}(\mathcal{X}_A\cap\mathcal{N}_{\mathrm{unt}}=\mathcal{M}_A
 \mid \mathcal{K}_{\text{unt}},\mathcal{N}_{\text{unt}})
 \nonumber \\ &= \Theta(\mathcal{M}_A,\mathcal{N}_{\rm unt}). 
 \end{align}
 In the DQPS protocol, Bob's basis choice can be postponed after he confirms photon detection, 
 which means that 
 the choice of $\mathcal{X}_B$ can be conducted after $\mathcal{K}_{\rm unt}$ and 
 $\mathcal{N}_{\rm unt}$ are determined. Hence, we have  
\begin{align}
 &\text{Pr}(\mathcal{X}_A\cap\mathcal{N}_{\mathrm{unt}}=\mathcal{M}_A,
 \mathcal{X}_B\cap\mathcal{N}_{\mathrm{unt}}=\mathcal{M}_B \mid \mathcal{K}_{\text{unt}},\mathcal{N}_{\text{unt}})
 \nonumber \\ &= \Theta(\mathcal{M}_A,\mathcal{N}_{\rm unt})\Theta(\mathcal{M}_B,\mathcal{N}_{\rm unt}),
\end{align}
which is identical to Eq.~(\ref{gozilla}). 
Similarly to the WCP-BB84 protocol, 
Eq.~(\ref{BIkunt}) 
holds, which leads to the Eq.~(\ref{fwcpb}): 
\begin{equation}
{\rm Pr}(k_{\rm ph,unt} > f_{\rm BI}(k_X))\leq \epsilon_{\rm PE}.\label{zola}
\end{equation}

The task of finding a bound $\munderbar{n}_{Z,\rm unt}$ satisfying \eq{kagen} is done as follows. 
In Ref.~\cite{2016Kawakami}, 
a modified protocol having exactly the same ${\rm Pr}(n_{Z,\rm tag})$ as the original protocol was introduced, 
in which a random variable $N$ 
(denoted as $n(c=d=0,(z_0'...z_{L-1}')\notin \Gamma^{(m)})$ in Eq.~(40) of 
Ref.~\cite{2016Kawakami}) 
satisfying $N\ge n_{Z,\rm tag}$ is defined. 
The variable obeys binomial distribution 
${\rm BI}(N,n_{\rm rep},r_{\rm tag} \tilde{p}_Z^2)$, 
where $r_{\rm tag}$ is a property of the light source defined as 
\begin{equation}
r_{\rm tag}:=1-\sum_m {\rm tr}(\hat{\Pi}_S^{(m)} \hat{\sigma}_S),
\label{tagtag}
\end{equation}
where $\hat{\sigma}_S$ is the state of $L$ pulses emitted from Alice's light source, 
and $\hat{\Pi}_S^{(m)}$ is a projector which is defined in Appendix. 
This implies that ${\rm Pr}(n_{Z,\rm tag})$ in the original protocol has the following property: 
There exists a function $P(n_{Z,\rm tag},N)$ satisfying 
\begin{align}
&{\rm Pr}(n_{Z,\rm tag})=\sum_N P(n_{Z,\rm tag},N)\nonumber \\
&P(n_{Z,\rm tag},N)=0~~{\rm for}~~n_{Z,\rm tag}>N \nonumber \\
&\sum_{n_{Z,\rm tag}}P(n_{Z,\rm tag},N)={\rm BI}(N,n_{\rm rep},r_{\rm tag} \tilde{p}_Z^2).
\label{torio}
\end{align}
This leads to 
\begin{equation}
{\rm Pr}(n_{Z,\rm tag}> n)\leq 1-C_{\rm BI}(n;n_{\rm rep},r_{\rm tag} \tilde{p}_Z^2)
\end{equation}
for any $n$, which is identical to Eq.~(\ref{hitori}). 
Then, following the same argument as the WCP-BB84 protocol, 
we see that 
\begin{align}
{\rm Pr}(n_{Z,\rm unt} < \munderbar{n}_{Z,\rm unt})\leq \epsilon_{Z,\rm unt}
\label{gonokami} 
\end{align}
holds 
with 
\begin{equation}
\munderbar{n}_{Z,\rm unt}:=n_Z-g(r_{\rm tag}\tilde{p}_Z^2,\epsilon_{Z,\rm unt}).
\label{gadgett}
\end{equation} 
%Although our tagging rule in the DQPS protocol is different from the one for the WCP-BB84 protocol, 
%the tagging process is still commuting with the actual key-generation process. 
%Thus, the secuirty parameter is set to be 
%\begin{equation}
%\epsilon_{\rm sec}=\epsilon_{\rm c}+\sqrt{2}\sqrt{\epsilon_{\rm PE}+\epsilon_{\rm PA}}+\epsilon_{Z,\rm unt}. 
%\end{equation}

From Eqs.~(\ref{keytag}), (\ref{zola}) and (\ref{gonokami}), 
we arrive at a key rate formula which is identical to \eq{keywcpb}: 
The $L$-pulse DQPS protocol is 
$\epsilon_{\rm c}$-correct and $\epsilon_{\rm s}$-secret if the final key length $n_{\rm fin}$ satisfies 
\begin{align} 
n_{\rm fin} \leq l_{\rm DQPS}
:=
\munderbar{n}_{Z,\rm unt}(1-h\left (\frac{f_{\rm BI}(k_X)}
{\munderbar{n}_{Z,\rm unt}}\right)) 
-{\rm log_2} \frac{2}{\epsilon_{\rm PA}} - \lambda_{\rm EC}(\epsilon_{\rm c})
\label{wasshoi}
\end{align} 
where 
$\epsilon_{\rm s}$ is 
given in \eq{gapao}. 
Together with Eqs.~(\ref{fb}), (\ref{cb}), (\ref{g}), (\ref{tagtag}) and (\ref{gadgett}), 
Eq.~(\ref{wasshoi}) constitutes the main result of Sec \ref{appDQPS}.

\begin{figure}[t]
  \begin{center}
  \raisebox{0mm}
  {\includegraphics[width=80mm]{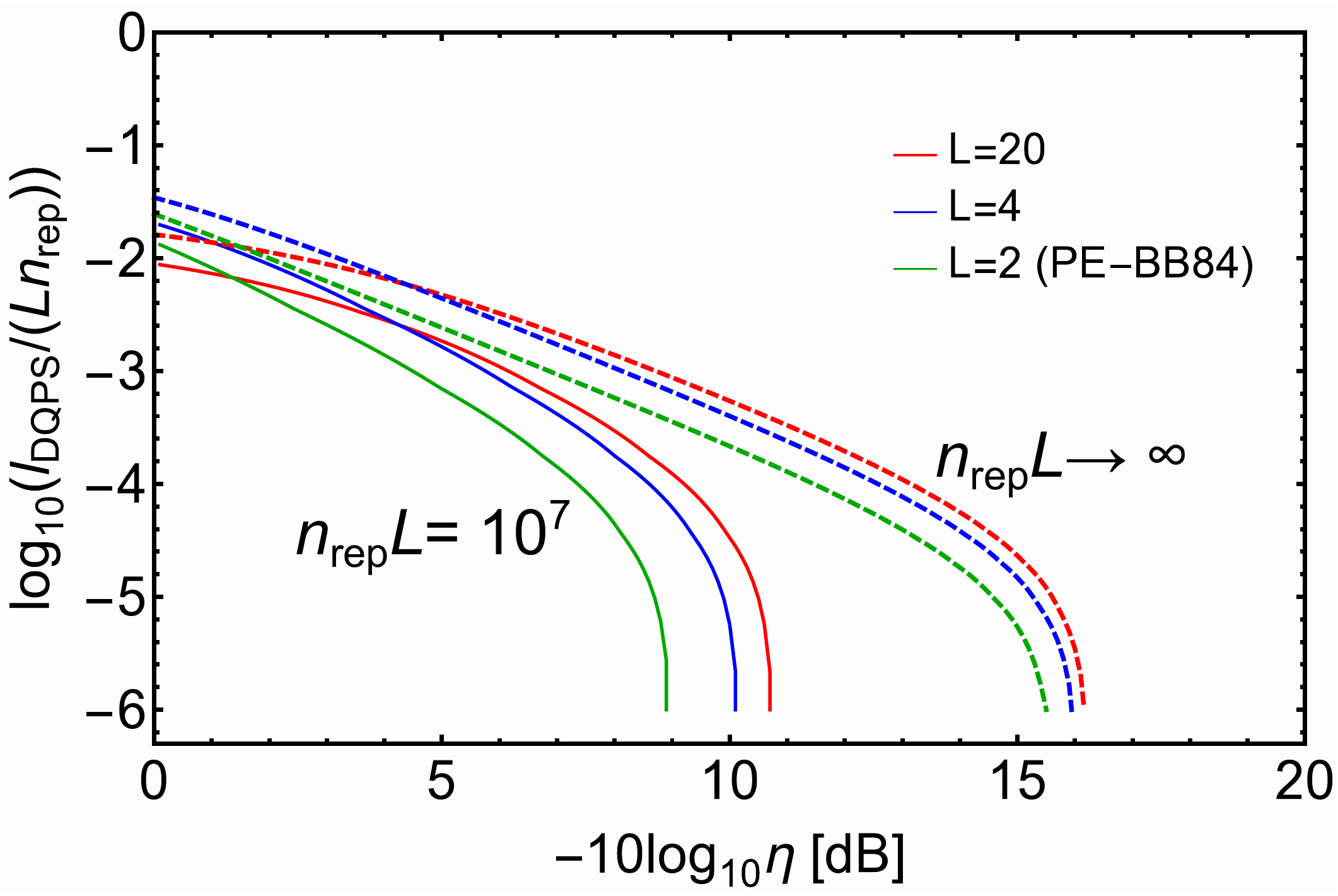}}
   \end{center}
   \caption{Secure key rate per pulse of the DQPS protocol 
   $l_{\rm DQPS}/(n_{\rm rep}L)$ as a function of 
   overall transmission $\eta$. 
   Solid curves are the results of the finite key analysis with total pulse number 
   $n_{\rm rep}L=10^{7}$ and dashed curves are the results of the asymptotic case ($n_{\rm rep}L\to \infty$), 
   which are obtained in Ref.~\cite{2016Kawakami}.   
   For both solid and dotted curves, the top, middle and bottom curves represent the key rate 
   for $L=20$, $L=4$ and $L=2$, respectively. The parameters are set as follows. 
Dark count rate per pulse per detector: $p_{\rm dark}=0.5\times 10^{-5}$. 
Loss-independent bit error: $3\%$. 
Cost for error correction: $\lambda_{\rm EC}(\epsilon_{\rm c})=1.1 h(E)+{\rm log}_2 (1/\epsilon_{\rm c})$. 
The security parameter: $\epsilon_{\rm c}=10^{-15}$ and $\epsilon_{\rm s}=10^{-10}$. 
We see that the key rate of the DQPS protocol $(L>2)$ is higher than that of the PE-BB84 protocol $(L=2)$ 
for both the asymptotic and finite-key cases. 
   }
   \label{DQPSkeyrate}
\end{figure}

In Fig.~\ref{DQPSkeyrate}, we show numerical results of 
 secure key rate per pulse $l_{\rm DQPS}/(n_{\rm rep}L)$ as a function of 
overall transmittance $\eta:= \eta_c \eta_d$ to compare the 
DQPS protocol ($L>2$) and the PE-BB84 protocol ($L=2$). 
The solid curves represent the key rate with fixed pulse number 
$n_{\rm rep}L=10^7$, and the dashed curves represent the one for the asymptotic case, 
which is obtained in our previous work \cite{2016Kawakami}. 
We assumed that Alice generates a weak coherent pulse of mean photon number $\mu$. 
In this case, 
 $r_{\rm tag}$ is given by 
\begin{equation}
r_{\rm tag}= 1- \sum_{m=0}^{\lceil L/2 \rceil} 
e^{-\mu L}\mu^m 
{}_{L+1-m}C_m.  
\label{rozario}
 \end{equation} 
 Note that for $L=2$, $r_{\rm tag}=1-e^{-2\mu}-2\mu e^{-2\mu}$ is 
 identical to the probability that two or more photons are emitted in a double-pulse signal in  
the PE-BB84 protocol. 
We assume dark count rate per pulse per detector 
$p_{\rm dark}=0.5\times 10^{-5}$ and a loss-independent bit error rate 3\%. 
We also assumed that 
$Q=1-(1-2(L-1)p_{\rm dark})e^{-(L-1)\mu \eta}$, reflecting the fact that there are $L-1$ valid 
timings in a block. 
Error rate is given by $E/Q$ where $E=0.03(1-e^{-(L-1)\mu \eta}) +p_{\rm dark}e^{-(L-1)\mu \eta} (L-1)$. 
Based on these parameters, we assume 
$\lambda_{\rm EC}(\epsilon_{\rm c})=1.1 h(E/Q)+{\rm log}_2 (1/\epsilon_{\rm c})$, $n_Z=n_{\rm rep}Q\tilde{p}_Z^2$, 
$n_X=n_{\rm rep}Q\tilde{p}_X^2$ and $k_X=n_X E/Q$. 
The values of $\tilde{p}_X$ and $\mu$ are optimized to maximize the key length. 
In the asymptotic limit, 
the parameter optimization leads to 
$\tilde{p}_X\to 0$, 
$\munderbar{n}_{Z,\rm unt}\to n_{\rm rep}(Q-r_{\rm tag})$ and  
$f_{\rm BI}(k_X)/\munderbar{n}_{Z,\rm unt} \to E/(Q-r_{\rm tag})$ 
while $Q$ and $E$ are fixed. 
In finite-key cases, 
the Chernoff bound is used to calculate the key rate. 
The security parameters are set to be the same as those in Fig.~\ref{wcpkeyrate}.
%$\epsilon_{\rm c}=10^{-15}$ and 
%$\epsilon_{\rm s}=10^{-10}$ 
%with $\epsilon_{\rm PE}=1/16\times 10^{-20}$, 
%$\epsilon_{\rm PA}=1/16\times 10^{-20}$ and $\epsilon_{Z,\rm unt}=1/2\times 10^{-10}$.  
We see that the advantage of the DQPS protocol over the PE-BB84 protocol 
is maintained even if we include the effect of the finiteness of the key.

\section{Concluding remarks}\label{conclusion}
In this paper, we proposed a method of finite-key analysis based on Bernoulli sampling 
instead of simple random sampling. 
For the BB84 protocol using biased basis choice, 
the data gathered on one of the basis is solely used for estimation of the disturbance in the other basis, 
which enables us to regard the former as a sample drawn from the population via Bernoulli sampling. 
As a result, we obtained finite-sized key-length formulas based on the binomial distribution 
parametrized by the probability of the basis choice in the protocol. 
The appearance of the binomial distribution in our case is a direct consequence of the 
inherent statistics of the protocol, and it should be differentiated from 
the previous works which uses a binomial distribution to derive an upper bound 
on the hypergeometric distribution arising from simple random sampling. 

The new method is particularly suited for the BB84 protocol with WCP. 
It enables simpler analysis compared to the method with simple random sampling 
since 
only the latter requires the estimation of the sample size ($n_{X,\rm unt}$). 
We may expect that this additional pessimistic bound makes the conventional method less efficient, 
which is corroborated by a numerical example showing that 
the key rate for the WCP-BB84 protocol obtained with our method is higher than that with simple random sampling. 
To make comparison with the previous finite-key analysis for the WCP-BB84 protocol~\cite{2009Scarani}, 
we calculated the key rate as a function of channel transmission and the number of detected signals, 
in the same practical parameter settings. The result shows that, while 
$n_{\rm det}\sim 10^7$ signals are necessary for producing a key in Ref.~\cite{2009Scarani}, 
our method only needs $n_{\rm det}\sim 10^4$ with the same parameters. 
In addition, the improved number $10^4$ clarifies that the use of WCP instead of 
an ideal single photon causes only a small change in the finite-size effect. 
This was also confirmed in the numerical simulation assuming the perfect channel, 
in which the required number of rounds to generate a key is $n_{\rm rep}\sim 10^{3.7}$ for the
 WCP-BB84 protocol and is $n_{\rm rep}\sim 10^{3.2}$ for the single-photon BB84 protocol.

Finally, we applied the Bernoulli-sampling method to the DQPS protocol, which was recently 
proved to be secure in the asymptotic regime. 
Although the asymptotic proof is based on the tagging of the insecure rounds as in the WCP-BB84 protocol, 
the definition of the tagged round is much more convoluted and makes sense only after the signal 
was detected by Bob. 
Nonetheless, our finite-key analysis has led to a key rate formula closely analogous to the one for 
the WCP-BB84 protocol. 
Numerical calculation 
shows that the DQPS protocol retains higher key rates than the BB84 protocol with phase encoding 
(PE-BB84) even in the finite-key regime of $n_{\rm rep}=10^7$.

It is expected that our method can also be applied to protocols with decoy states~\cite{2003Hwang,2005WangPRL,2005Decoy}. 
Since the existing analyses~\cite{2014Wen,2014Curty,2014Hayashi,2015Lucamarini} with decoy states involve the estimation of 
the sample size $n_{X,\rm unt}$, 
the present method may provide a simpler analysis compared to the conventional methods 
with simple random sampling. 
It should be mentioned that some of the finite key analyses~\cite{2014Wen,2014Curty} 
assumed the announcement of basis choice after each round to make the sample size fixed, 
which were later pointed out~\cite{2016Pfister} to open a security hole against a sifting attack. 
This illustrates an importance of simpler and more straightforward methods, 
and we believe that the method proposed here will contribute in this regard.

\section*{Acknowledgement}
We thank A. Mizutani, T. Tsurumaru and K. Yoshino for helpful discussions. 
This work was
supported by the ImPACT Program of the Council for Science, Technology and
Innovation (Cabinet Office, Government of Japan), CREST, and the Photon Frontier Network 
Program (MEXT).

\appendix*

\section{Description of the DQPS protocol}
Here we summarize the detail of the DQPS protocol in Ref.~\cite{2016Kawakami}. 
%The protocol uses  two bases, data basis for generating the final key  and check basis 
%for monitoring the leak of information. 
%In the data and check bases, relative phases between adjacent pulses are modulated by $\{0,\pi\}$ and 
%$\{\frac{\pi}{2},\frac{3 \pi}{2}\}$, respectively.
%The protocol regards a train of $L$ pulses as a block, 
%and the working basis is randomly chosen for each block. 
%The randomization of overall optical phase is also done for each block of $L$ pulses. 
%Bob's receiver is composed of delayed interferometer with its delay being equal to 
%the interval $\Delta{\tau}$ of adjacent pulses.  
%The longer arm of the interferometer passes through a phase modulator that incurs phase shift 
%$\theta_B=$ 0 or $\frac{\pi}{2}$. 
The protocol proceeds as follows, 
which includes predetermined parameters $\tilde{p}_X>0$,  $\tilde{p}_Z=1-\tilde{p}_X$, $L\geq 2$, and $n_{\rm rep}$. 
\\
1.~~Alice selects a bit $c\in \{0,1\}$ with probability $\tilde{p}_Z$ and $\tilde{p}_X$, 
which correspond to the choice of $Z$ basis and $X$ basis, respectively. 
Bob also selects $d\in \{0,1\}$ with probability $\tilde{p}_Z$ and $\tilde{p}_X$.\\
2.~~Alice generates $L$ random bits $a_l\in \{0,1\}$ $(l=0,1,..,L-1)$, and 
prepares 
$L$ optical pulses (system $S$) in the state
\begin{align}
\hat{\rho}_S&=\hat{S}(\{a_l\},c)\hat{\sigma}_S\hat{S}(\{a_l\},c) \nonumber \\
\hat{S}(\{a_l\},c)&:=\bigotimes_{l=0}^{L-1}{\rm exp}(i (a_l \pi +\frac{\pi}{2}lc)\hat{n}_l)
\label{pre}
\end{align}
where $\hat{\sigma}_S$ is the state of the $L$ pulses from the source before phase modulation and 
$\hat{n}_l$ represents the photon number operator for the $l$-th pulse.  
Alice randomizes the overall optical phase of the $L$-pulse train, and sends it to Bob. 
\\
3.~~If $d=0$, Bob sets the phase shift $\theta_B=0$. If $d=1$, he sets $\theta_B=\frac{\pi}{2}$.\\
4.~~If there is no detection of photons at the valid timings, Bob sets $j=0$. 
If the detections have only occurred at a single valid timing, 
the variable $j$ is set to the index of the timing.
If there are detections at multiple timings,
the smallest (earliest) index of them is assigned to $j$. 
If $j\neq 0$, Bob determines his raw key bit $b\in \{0,1\}$ depending on which detector has reported detection at the 
$j$-th timing. 
If both detectors have reported at the $j$-th timing, a random bit is assigned to $b$.
Bob announces $j$ publicly.\\ 
5.~~If $j\neq 0$, Alice determines her raw key bit as $a=a_{j-1}\oplus a_j$.\\
6.~~Alice and Bob repeat the above procedures $n_{\rm rep}$ times. 
They publicly disclose $c$ and $d$ for each of the $n_{\rm rep}$ rounds. \\
7.~~Alice and Bob define bit strings  
$\bm{\kappa}_{A,X}$ and 
$\bm{\kappa}_{B,X}$, respectively, by concatenating their determined bits with $j\neq 0$ and $c=d=1$. 
They define sifted keys 
$\bm{\kappa}_{A,Z}$ and $\bm{\kappa}_{B,Z}$, respectively, 
by concatenating their determined bits with $j\neq 0$ and $c=d=0$. 
Let their sizes be $n_Z:=|\bm{\kappa}_{A,Z}|=|\bm{\kappa}_{B,Z}|$ and 
$n_X:=|\bm{\kappa}_{A,X}|=|\bm{\kappa}_{B,X}|$.\\
8.~~They disclose and compare $\bm{\kappa}_{A,X}$ and $\bm{\kappa}_{B,X}$ to determine the number of bit errors $k_X$ 
included in them.\\ 
9.~~Through public discussion, 
Bob corrects his keys $\bm{\kappa}_{B, Z}$ to make it coincide 
with Alice's key $\bm{\kappa}_{A, Z}$ and obtains $\bm{\kappa}_{B, Z}^{\rm cor}$ 
$(|\bm{\kappa}_{B, Z}^{\rm cor}|=n_Z)$. \\
10.~~Alice and Bob conduct privacy amplification by shortening $\bm{\kappa}_{A,Z}$ and 
$\bm{\kappa}_{B,Z}^{\rm cor}$ to obtain final keys $\bm{\kappa}_{A,Z}^{\rm fin}$ and 
$\bm{\kappa}_{B,Z}^{\rm fin}$ 
of size $n_{\rm fin}$. \\

The security of the above protocol in the asymptotic limit was proved in \cite{2016Kawakami} 
under the following 
assumptions on the devices used by Alice and Bob. 
The discussion in subsection \ref{appDQPS} of the main text uses the same assumptions. 
We assume that the phase randomization in Step 2 is ideal, 
and hence the state emitted from Alice in Step 2 is expressed as 
\begin{equation}
\sum_{m} \hat{N}_m \hat{\rho}_S\hat{N}_m, 
\label{enuemu}
\end{equation}
where $\hat{N}_m$ represents the projector onto the subspace with $m$ total photons in the $L$ pulses. 
We also assume that a parameter $r_{\rm tag}$ associated with the $L$-pulse state 
$\hat{\sigma}_S$ from the source is known or at least is bounded from above. 
With $\ket{m_l}_{S,l}$ being an $m$-photon state of the $l$-th pulse, the parameter is defined by
\eq{rozario} with
\begin{equation}
\hat{\Pi}^{(m)}_{S}:= \sum_{\{m_l\}\in \Gamma^{(m)}} \bigotimes_{l=0}^{L-1}
  \ket{m_l}\bra{m_l}_{S,l}
\end{equation}
where 
$\Gamma^{(m)}$ is 
 a set of values of $L$ nonnegative integers 
\begin{equation}
 \Gamma^{(m)}:=\left\{(i_0,\cdots, i_{L-1}) ~\Bigg |~  i_{l-1}+i_l \le
  1 (1\le l \le L-1), \sum_{l=0}^{L-1} i_l=m \right\}.
\end{equation}
In \cite{2016Kawakami}, we showed a practical method of off-line calibration  
to determine an upper bound of $r_{\rm tag}$ for a general light source.

To describe the assumptions for Bob's apparatus, 
we introduce POVM elements for Bob's procedure in Steps 3 and 4. Let 
$\{\hat{B}^{(d)}_j\}_{j=0,...,L-1}$ be the POVM 
for Bob's procedure of determining $j$, when the basis $d$ was selected in Step 1.  
We further decompose the elements for $j\neq 0$ as $\hat{B}^{(d)}_j=\hat{B}^{(d)}_{j,0} + \hat{B}^{(d)}_{j,1}$, where 
$\hat{B}^{(d)}_{j,b}$ corresponds to the outcome $(j,b)$. These operators satisfy
\begin{equation}
\hat{B}^{(d)}_0+\sum_{j=1}^{L-1} 
(\hat{B}^{(d)}_{j,0} + \hat{B}^{(d)}_{j,1}) = \hat{\mathbbm{1}}.
\end{equation}
We then 
assume that Bob uses threshold detectors, and further assume that 
 the inefficiency and dark countings of the detectors are 
equivalently represented by an absorber and a stray photon source placed in front of Bob's 
apparatus, and hence they are included in the quantum channel.  
This leads \cite{2016Kawakami} to the condition 
\begin{equation}
\hat{B}_j^{(0)}=\hat{B}_j^{(1)}~~~(0\leq j\leq L-1). 
\label{honnnetotatemae}
\end{equation}


%\bibliographystyle{apsrev4-1}
%%\bibliographystyle{unsrt}
%%\bibliographystyle{iopart-num}
%%\bibliographystyle{mymastersty_etal}
%\bibliography{sasakibib,NJPyoubib,kawakamibib,kawakamibibD}

\end{document}